\documentclass[instruments,article,accept,moreauthors,pdftex]{Definitions/mdpi} 

\firstpage{1} 
\pubvolume{4}
\issuenum{3}
\articlenumber{21}
\pubyear{2020}
\copyrightyear{2020}
\history{Received: 29 June 2020; Accepted: 24 July 2020; Published: 28 July 2020}

\setitemize{parsep=6pt,itemsep=0pt,leftmargin=*,labelsep=5.5mm}
\setenumerate{parsep=6pt,itemsep=0pt,leftmargin=*,labelsep=5.5mm}
\setlist[description]{itemsep=0mm}

\usepackage{amsmath}           
\usepackage{amssymb}
\usepackage[UKenglish]{babel}
\usepackage{bm}                
\usepackage{booktabs}
\usepackage{color}
\usepackage{enumerate}
\usepackage{fancyhdr}
\usepackage{float}
\usepackage{graphicx}
\usepackage[space]{grffile}    
\usepackage[utf8]{inputenc}
\usepackage{longtable}
\usepackage{marvosym}
\usepackage{multicol}
\usepackage[separate-uncertainty,multi-part-units = single, allow-number-unit-breaks]{siunitx}
\usepackage{subfig}
\usepackage{tikz-timing}
\usepackage[colorinlistoftodos,prependcaption,textsize=tiny]{todonotes} 

\usepackage{url}
\def\UrlBreaks{\do\/\do-}

\newcommand*\patchAmsMathEnvironmentForLineno[1]{%
  \expandafter\let\csname old#1\expandafter\endcsname\csname #1\endcsname
  \expandafter\let\csname oldend#1\expandafter\endcsname\csname end#1\endcsname
  \renewenvironment{#1}%
  	{\linenomath\csname old#1\endcsname}%
	{\csname oldend#1\endcsname\endlinenomath}}%
\newcommand*\patchBothAmsMathEnvironmentsForLineno[1]{%
  \patchAmsMathEnvironmentForLineno{#1}%
  \patchAmsMathEnvironmentForLineno{#1*}}%
\AtBeginDocument{%
\patchBothAmsMathEnvironmentsForLineno{equation}%
\patchBothAmsMathEnvironmentsForLineno{align}%
\patchBothAmsMathEnvironmentsForLineno{flalign}%
\patchBothAmsMathEnvironmentsForLineno{alignat}%
\patchBothAmsMathEnvironmentsForLineno{gather}%
\patchBothAmsMathEnvironmentsForLineno{multline}%
}

\usepackage{adjustbox}

\Title{Off-Axis Characterisation of the CERN T10 Beam for low Momentum Proton Measurements with a High Pressure Gas Time Projection Chamber}

\Author{
  {S.B.~Jones} 
  { $^{1,}$}*\orcidA{}, 
  T.S.~Nonnenmacher $^{2,}$*,    
  E.~Atkin $^{2}$, 
  G.J.~Barker $^{3}$, 
  A.~Basharina-Freshville $^{1}$,
  C.~Betancourt $^{4}$,
  S.B.~Boyd $^{3}$, 
  D.~Brailsford $^{5}$, 
  Z.~Chen-Wishart $^{6}$, 
  L.~Cremonesi $^{1}$, 
  A.~Deisting $^{6}$,  
  A.~Dias $^{6}$,
  P.~Dunne $^{2}$,  
  J.~Haigh $^{3}$, 
  P.~Hamacher-Baumann $^{7}$, 
  A.~Kaboth $^{6}$,   
  A.~Korzenev $^{8}$,
  W.~Ma $^{7}$,
  P.~Mermod $^{8}$,
  M.~Mironova $^{2,9}$,
  J.~Monroe $^{6}$, 
  R.~Nichol $^{1}$, 
  J.~Nowak $^{5}$, 
  W.~Parker $^{6}$, 
  H.~Ritchie-Yates~$^{6}$,
  S.~Roth~$^{7}$,
  R.~Saakyan $^{1}$, 
  N.~Serra $^{4}$,
  Y.~Shitov $^{2,10}$,  
  J.~Steinmann $^{7}$, 
  A.~Tarrant $^{6}$, 
  M.A.~Uchida $^{2,11}$,  
  S.~Valder $^{3}$, 
  A.V.~Waldron $^{2}$, 
  M.~Ward $^{6,12}$ and
  M.O.~Wascko $^{2}$
}

\AuthorNames{Sebastian Jones, Toby Nonnenmacher, Edward Atkin, Gary Barker, Anastasia Basharina Freshville, Christopher Betancourt, Steve Boyd, Dominic Brailsford, Zachary Chen-Wishart, Linda Cremonesi, Alexander Deisting, Adriana Dias, Patrick Dunne, Jennifer Haigh, Philip Hamacher-Baumann, Asher Kaboth, Alexander Korzenev, William Ma, Philippe Mermod, Maria Mironova, Jocelyn Monroe, Ryan Nichol, Jaroslaw Nowak, William Parker, Harrison Ritchie-Yates, Stefan Roth, Ruben Saakyan, Nicola Serra, Yuri Shitov, Jochen Steinmann, Adam Tarrant, Melissa Uchida, Sammy Valder, Abigail Waldron, Mark Ward and Morgan Wascko}

\address{%
        $^{1}$ \quad Department of Physics and Astronomy, University College London, Gower St, Kings Cross, \mbox{London WC1E 6BT}, UK; anastasia.freshville@ucl.ac.uk~(A.B.-F.); l.cremonesi@ucl.ac.uk (L.C.); r.nichol@ucl.ac.uk~(R.N.); r.saakyan@ucl.ac.uk~(R.S.)

        $^{2}$ \quad The Blackett Laboratory, Imperial College London, London SW7 2BW, UK; e.atkin17@imperial.ac.uk (E.A.); p.dunne12@imperial.ac.uk (P.D.); maria.mironova@physics.ox.ac.uk (M.M.); Shitov@JINR.ru (Y.S.); m.a.uchida@imperial.ac.uk (M.A.U.); a.waldron@imperial.ac.uk (A.V.W.); \mbox{m.wascko@imperial.ac.uk (M.O.W.)}

	$^{3}$ \quad Department of Physics, University of Warwick, Coventry CV4 7AL, UK; g.j.barker@warwick.ac.uk (G.J.B.); S.B.Boyd@warwick.ac.uk (S.B.B.); J.Haigh.2@warwick.ac.uk (J.H.); s.valder@warwick.ac.uk (S.V.)

        $^{4}$ \quad Physik-Institut, Universit\"at Z\"uriche, R\"amistrasse 71, 8006 Z\"urich, Switzerland; christopher.betancourt@cern.ch (C.B.); nicola.serra@cern.ch (N.S.)

	$^{5}$ \quad Department of Physics, Lancaster University, Bailrigg, Lancaster LA1 4YW, UK; \mbox{d.brailsford@lancaster.ac.uk (D.B.)}; j.nowak@lancaster.ac.uk (J.N.)

	$^{6}$ \quad Department of Physics, Royal Holloway, University of London, Egham Hill, Egham TW20 0EX, UK; Zachary.Chen-Wishart.2016@live.rhul.ac.uk (Z.C.-W.); Alexander.Deisting@cern.ch (A.D.); Adriana.Dias.2011@live.rhul.ac.uk (A.D.); Asher.Kaboth@rhul.ac.uk (A.K.); jocelyn.monroe@rhul.ac.uk~(J.M.); William.Parker.2016@live.rhul.ac.uk (W.P.); Harrison.Ritchie-Yates.2013@live.rhul.ac.uk (H.R.-Y.); Adam.Tarrant.2015@live.rhul.ac.uk (A.T.); mark.ward@snolab.ca (M.W.)

  	$^{7}$ \quad III. Physikalisches Institut,  RWTH Aachen University, 52056 Aachen, Germany; hamacher.baumann@physik.rwth-aachen.de (P.H.-B.); ma@physik.rwth-aachen.de (W.M.); stefan.roth@physik.rwth-aachen.de (S.R.); jochen.steinmann@physik.rwth-aachen.de (J.S.)

        $^{8}$ \quad DPNC Universit\'e de Gen\`eve, 1205 Genf, Switzerland;  korzenev@mail.cern.ch (A.K.); Philippe.Mermod@cern.ch (P.M.)

 	$^{9}$ \quad Department of Physics, Oxford University, Oxford OX1 3PU, UK

	$^{10}$ \quad JINR, 141980 Dubna, Russia
 
	$^{11}$ \quad Cavendish Laboratory, Cambridge CB3 0HE, UK
		
        $^{12}$ \quad Department of Physics, Queen’s University, Kingston, ON K7L 3N6, Canada

}

\corres{Correspondence: sebastian.jones.17@ucl.ac.uk (S.B.J.); toby.nonnenmacher14@imperial.ac.uk (T.S.N.)}

\abstract{We present studies of proton fluxes in the T10 beamline at CERN.
A prototype high pressure gas time projection chamber (TPC) was exposed to the beam of protons and other particles, using the 0.8~GeV/c momentum setting in T10, in order to make cross section measurements of low energy protons in argon.
To explore the energy region comparable to hadrons produced by GeV-scale neutrino interactions at oscillation experiments, i.e., near 0.1~GeV of~kinetic energy, methods of moderating the T10 beam were employed:
the dual technique of moderating the beam with acrylic blocks and measuring scattered protons off the beam axis was used to decrease the kinetic energy of~incident protons, as well as change the proton/minimum ionising particle (MIP) composition of the incident flux.
Measurements of the beam properties were made using time of flight systems upstream and downstream of the TPC.
The kinetic energy of protons reaching the TPC was successfully changed from $\sim$0.3~GeV without moderator blocks to less than 0.1~GeV with four moderator blocks (40~cm path length).
The flux of both protons and MIPs off the beam axis was increased.
The ratio of protons to MIPs vary as a function of the off-axis angle allowing for possible optimisation of the detector to select the type of required particles. 
Simulation informed by the time of flight measurements show that with four moderator blocks placed in the beamline,  ($5.6 \pm 0.1$) protons with energies below 0.1~GeV per spill traversed the active TPC region.
Measurements of the beam composition and energy are presented.}


\begin{document}

\section{Introduction}

One of the major goals of the global neutrino physics programme is to explore fundamental symmetries of nature linked to why we live in a matter-dominated universe. 
Charge-parity symmetry violation (CPV) in the neutrino sector is one possibility remaining to be explored further experimentally, and neutrino experiments strive to improve current measurements of CPV in the leptonic sector~\cite{Abe:2019vii}. 
CPV is obtained from the simultaneous fit of the $\nu_{\mu}$ disappearance and $\nu_e$ appearance oscillation channels separately for neutrinos and anti-neutrinos.
In the absence of CPV and accounting for matter effects, the rates of $\nu_{\mu}\!\rightarrow\!\nu_e$ and $\overline{\nu}_{\mu}\!\rightarrow\!\overline{\nu}_e$ oscillations should be equal.
To convert the measured rate of interactions to a level of CPV, experiments must accurately know the cross section for the interactions of neutrinos and anti-neutrinos with detector materials, which are most commonly hydrogen, carbon, oxygen, argon and iron. 
Therefore, systematic uncertainties on neutrino--nucleus interaction cross sections are a key input to such CPV searches.  
These interaction cross sections are dependent on modelling neutrino-nucleon interactions occurring within nuclei. 

The nuclear models informed by these cross sections have substantial effects on the measured final-state particle kinematic distributions~\cite{Mosel:2016cwa}.

The long baseline neutrino experiments that are currently searching for CPV are the Tokai to Kamioka experiment~(T2K)~\cite{Abe:2019vii} and the NuMI Off-Axis $\nu_{e}$ Appearance experiment~(NOvA)~\cite{Acero:2019ksn}.
The T2K experiment, which currently reports the strongest constraint on CPV in neutrinos~\cite{Abe:2019vii}, has systematic uncertainties of {7}--9\% ~
after near-detector constraint on the prediction of the rate of far detector electron-like events, with cross section uncertainties being the largest contribution.
The~future Deep Underground Neutrino Experiment~(DUNE)~\cite{abi2020deep} and Hyper-Kamiokande~\cite{abe2011letter} projects will seek to reach {1}--3\%~
on that same rate of far detector electron-like events~\cite{acciarri2016long}, with improved systematic errors providing better precision on the CP violating phase.
The key to reducing these uncertainties is to precisely measure the multiplicity and momentum distribution of final-state particles. 
\mbox{However, these distributions} are modified by final state interactions (FSI) of the recoiling secondary particles as they traverse the target nucleus. 
The most commonly used neutrino generator Monte Carlos (GENIE~\cite{Andreopoulos:2009rq}, NEUT~\cite{Hayato:2009zz} and NuWro~\cite{GOLAN2012499}), simulate FSI with cascade models that are tuned with external hadron--nucleus scattering measurements. The generator GiBUU~\cite{lalakulich2013neutrino} models FSI by solving the semi-classical Boltzmann--Uehling--Uhlenbeck equation.

\begin{figure}[H]
    \centering
    \includegraphics[width=12cm]{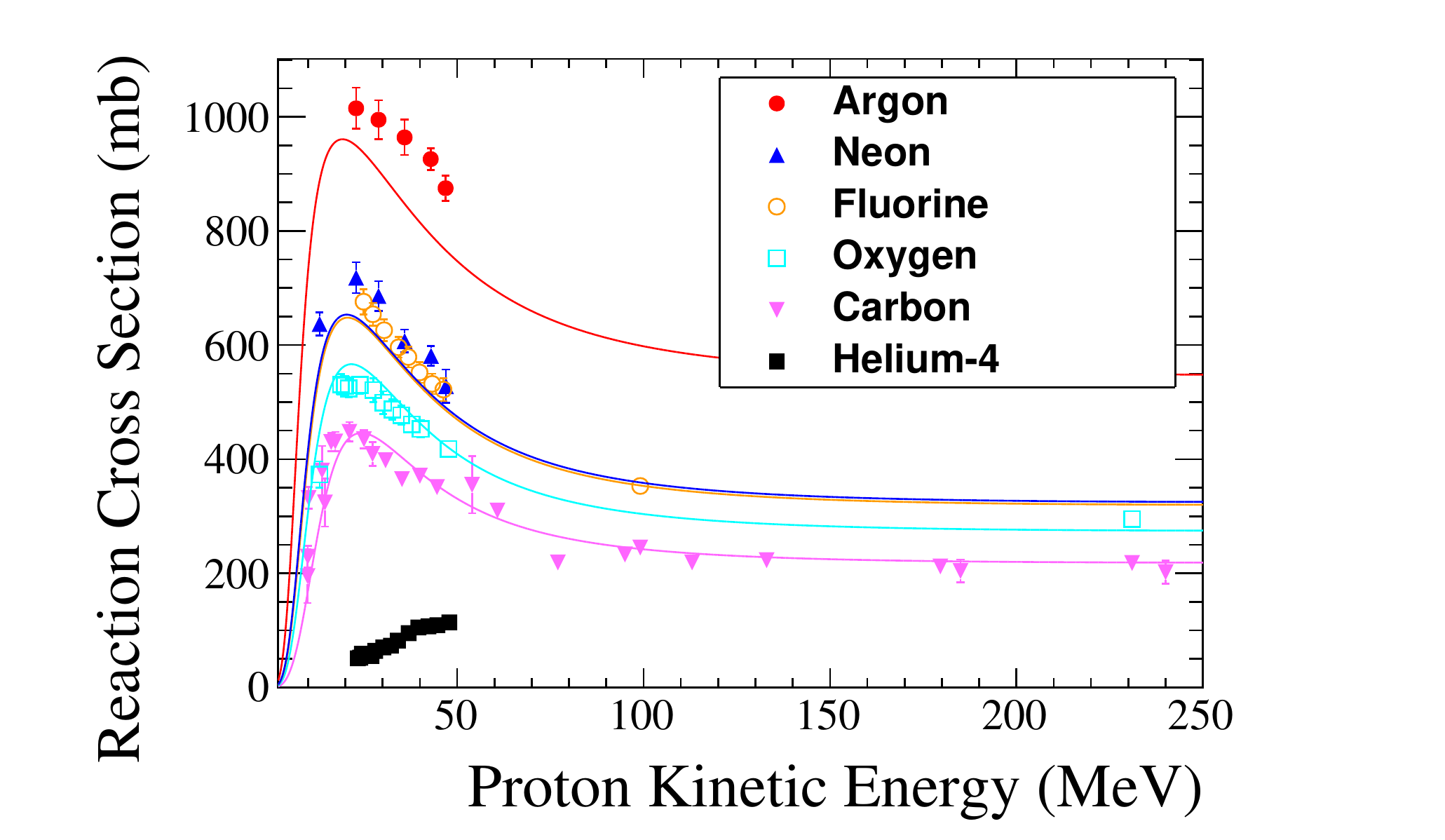}%
    \caption{Total reaction cross sections for protons on argon, neon, fluorine, oxygen, carbon and helium-4. {Data}~\cite{Carlson:1996ofz} are compared to a semi-empirical {model}~\cite{wellisch1996total}. Figure from~\cite{SPSC-P-355}.}
    \label{fig:DataProtonXSec}%
\end{figure}

However, as shown in Figure~\ref{fig:DataProtonXSec}, proton--nucleus scattering measurements are extremely sparse and in many cases do not exist in the relevant energy region and/or on the relevant nuclei.
Therefore semi-empirical parametrisations are used to extrapolate in momentum and atomic mass~\cite{wellisch1996total}.  
The~parametrisations are different between the three generators, and yield order-of-magnitude scale differences in the predicted multiplicity and kinematics of final state protons~\cite{dune2018high}.
The proton final state modelling is a key ingredient for neutrino oscillation measurements because it affects the event selection and neutrino energy reconstruction in charged-current (CC) interactions, which is the channel used to measure oscillation parameters and is therefore central to the search for CPV~\cite{Abe:2013hdq}.
For these reasons, FSI contribute substantially to the total neutrino interaction systematic uncertainty~\cite{Abe:2019vii}.

Moreover, FSI models are in tension with data.  
Recent neutrino scattering measurements have shown that the most-used models of neutrino-nucleus interactions (employed by NEUT and GENIE) differ from nature in both cross section and kinematics of final state particles by as much as 30\%~\cite{McFarland:2018aaa}. 
These uncertainties cannot be fully mitigated with near/far detector combinations because they come from theoretical model deficiencies that are not cancelled in the near–far extrapolation~\cite{Coloma:2013rqa}. 

\begin{figure}[H]
    \centering
    \includegraphics[width=12cm]{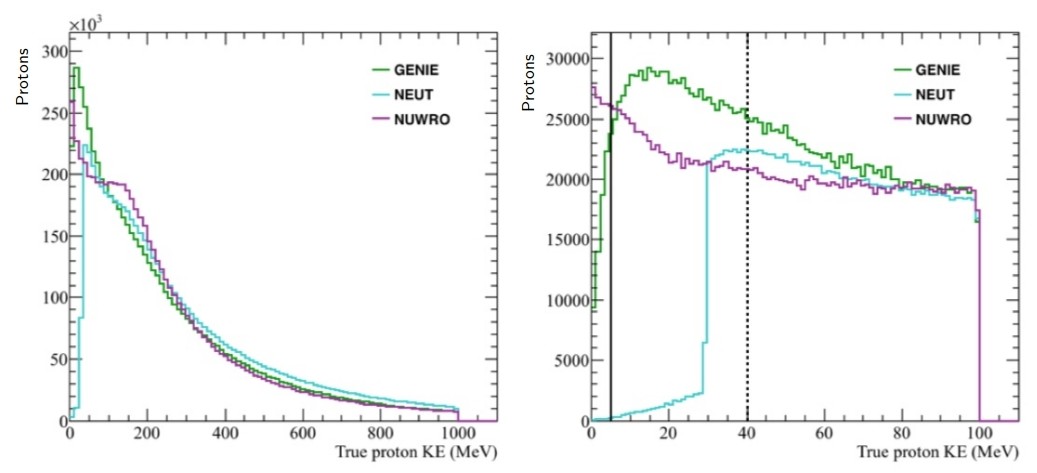}%
    \caption{{Predicted proton} kinetic energy (KE) spectra from GENIE, NEUT and NuWro~\cite{Raaf:2018aaa}. \mbox{Energy spectra} up to 1 GeV are shown on the left, and zoomed in to lower energies on the right. The figure uses the Long Baseline Neutrino Facility (LBNF)  simulation for DUNE's beam energy and flux. The LBNF beam has a mean energy of approximately 2.5~GeV~\cite{abi2020deep}. The dashed vertical line indicates the expected proton automated-reconstruction/identification threshold in liquid argon, and the solid vertical line shows the same for gaseous argon at 10 atm~\cite{dune2018high}.}
    \label{fig:protonsfromargon}%
\end{figure}

The key proton kinetic energy range in which to distinguish interaction models is the region below 0.1~GeV.
Figure~\ref{fig:protonsfromargon} shows the proton multiplicity and kinetic energy distributions for $\nu_{\mu}$ CC interactions on argon calculated by the GENIE, NEUT and NuWro neutrino generators for the DUNE experiment.
These distributions are highly discrepant at low proton kinetic energy as shown in the right hand panel. The generators are not designed to handle the low energy region consistently, due~to~the lack of available data.
This is predominantly below the proton detection threshold in liquid Argon TPCs (0.04~GeV), such as those that will be used by DUNE, and in water Cherenkov detectors (0.5~GeV).
\mbox{The lower} threshold in high pressure gas provides a unique opportunity to distinguish between neutrino interaction models for the same nuclear target.

\begin{figure}[H]
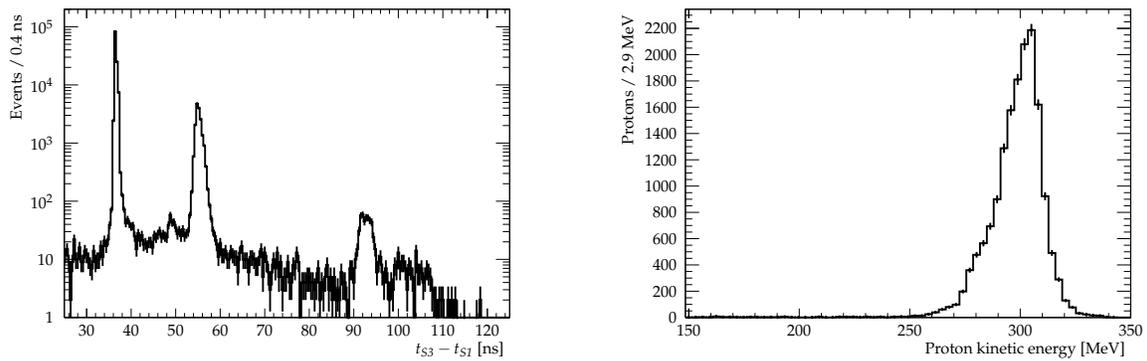

  \begin{minipage}[t]{0.49\textwidth}
    \centering
    \begin{adjustbox}{max totalsize={\textwidth},center}
      \input{files/Figures/utofNoBend.tex}
    \end{adjustbox}
  \end{minipage}
  \hfill
  \begin{minipage}[t]{0.49\textwidth}
    \centering
    \begin{adjustbox}{max totalsize={\textwidth},center}
      \input{files/Figures/protonKE.tex}
    \end{adjustbox}
  \end{minipage}
  \caption{\label{fig:utofNoBend}Measurements of the unmoderated and unbent T10 beam over a baseline of 10.8~m for \mbox{a selected} beam momentum of 0.8~GeV/c. Measurements are made in the $\mathit{S3}$ detector. The peak between 50~ns and 60~ns is produced by protons. (\textbf{Left}) Time of flight spectrum. (\textbf{Right}) Measured kinetic energy of protons.}
\end{figure}

We have built a High Pressure gas Time Projection Chamber (HPTPC) prototype and exposed it to a charged particle beam in the T10 beamline at CERN in August and September 2018 \cite{SPSC-P-355}.
The~momentum profile of the T10 beam can be tuned within the range 0.8--6.5~GeV/c (kinetic energy range 0.3--5.6~GeV). 
Figure~\ref{fig:utofNoBend} left, shows the time of flight (ToF) spectrum for the T10 beamline tuned to a momentum of 0.8~GeV/c; this measurement was made with our upstream ToF system (see Section~\ref{hptpcPaper:sec:Methods} for details of the ToF systems).
The kinetic energy of the protons calculated from the upstream ToF measurements in this sample is shown in Figure~\ref{fig:utofNoBend} right.
As shown, the flux of protons with kinetic energy less than 200~MeV is negligible.
The physics objective of the HPTPC beam test was to make measurements of protons on argon at kinetic energies below 200~MeV, i.e., below what was available with the T10 beam. 
Furthermore, the readout speed of the charge-coupled device cameras (CCD) employed in the HPTPC prototype motivates a limit on the total particle multiplicity in the TPC active~volume.

To enhance the low energy proton flux, a novel technique was employed:
we placed acrylic moderator blocks directly in the beamline, which spread and slowed the beam particles via multiple Coulomb scattering.
By placing the TPC in an off-axis position with respect to the beam direction, \mbox{we observed} a beam composition with lower-energy protons than would otherwise have been possible in the T10 beamline.
These techniques were designed to increase the ratio of protons to MIPs in the TPC, and to decrease the proton momentum and multiplicity in the active region of the TPC.

The flux and composition of beam particles were measured with two ToF systems, \mbox{placed upstream} and downstream of the TPC.
Measurements of protons and MIPs are presented as a function of the off-axis angle and thickness of the moderator.
This paper provides a detailed description of the time of flight systems employed in the beamline in Section~\ref{hptpcPaper:sec:Methods}, the analysis methodology of the ToF data in Section~\ref{hptpcPaper:sec:Analysis}, presentation of the ToF system results in Section~\ref{hptpcPaper:sec:Results} and additional conclusions in Section~\ref{hptpcPaper:sec:Conclusion}.

\section{Beam Line and Detectors}
\label{hptpcPaper:sec:Methods}
\vspace{-6pt}
\subsection{Beam Test Overview}
\begin{figure}[h]
  \centering
  \includegraphics[width=0.43\linewidth]{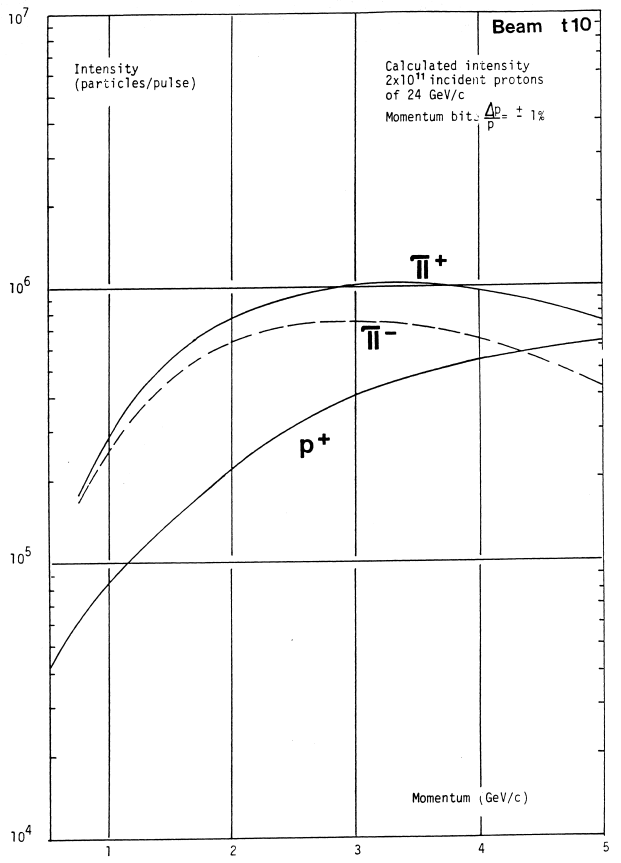}
  \caption{Calculated intensity of the T10 beam as a function of selected beam momentum, separated by particle type~\cite{T10Report}.}
  \label{fig:beamComp}
\end{figure}

\begin{figure}[H]
  \includegraphics[width=1.0\linewidth]{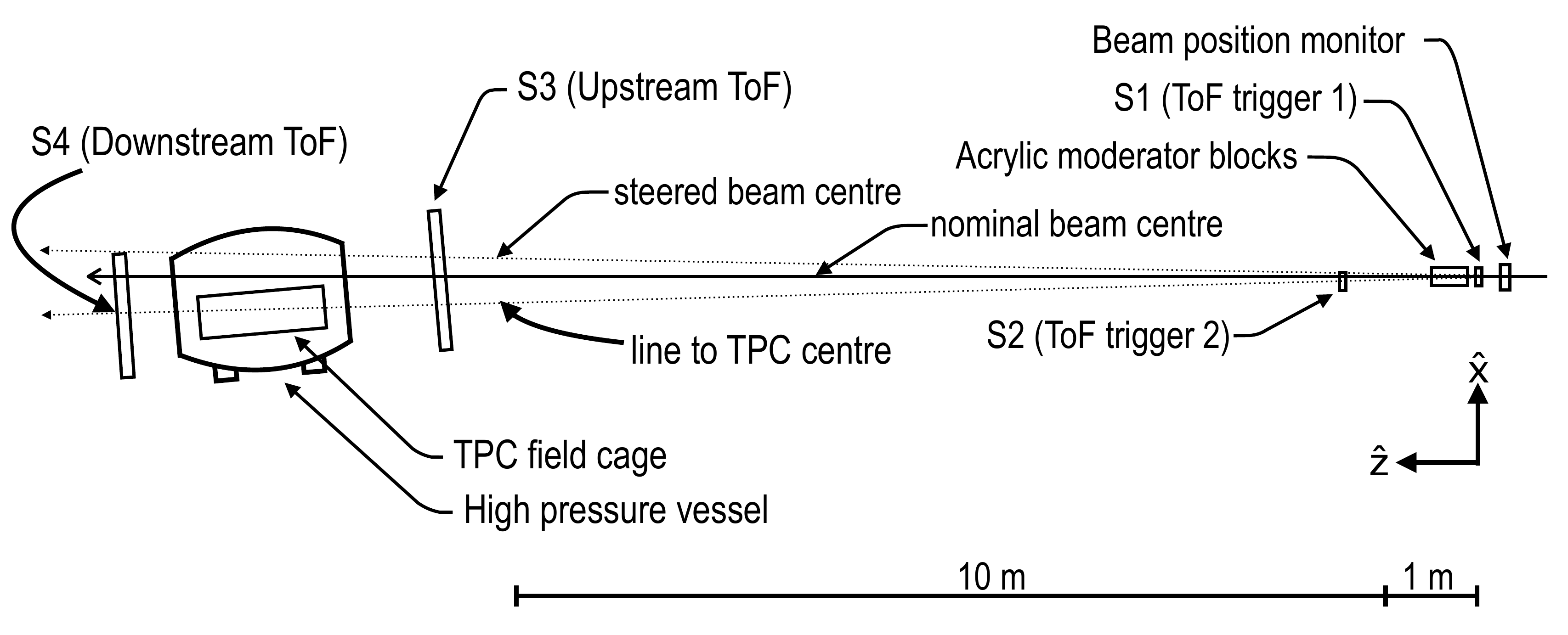}
\captionsetup{width=1\linewidth}
  \caption{Schematic diagram (plan view) of the  High Pressure gas Time Projection Chamber (HPTPC) beam test configuration in the T10 area at CERN.} 
  \label{fig:setup}
\end{figure}

The beam test took place in the T10 beam line, in the East Area at the Proton Synchrotron (PS) at CERN.
The T10 beamline at CERN is a secondary beam derived from the PS beam which consists primarily of protons, electrons and charged pions~\cite{T10Report}.
The theoretical beam composition as a function of beam momentum is shown in Figure~\ref{fig:beamComp}.
The primary components of the experimental setup are shown schematically in Figure~\ref{fig:setup}.

A beam position monitor (BPM) was situated at the beam entrance into the test area, upstream of all the ToF constituents and the TPC. 
The TPC was placed 13~m downstream of the BPM. 
\mbox{From initial} GEANT4~\cite{brun1993geant} beam simulations, the optimal TPC position to reduce the momentum of particles reaching the detector, without excessively reducing particle flux, was determined to be between 2$^{ \circ }$ and 3$^{ \circ }$ off the beam axis, but space constraints meant the TPC could not be placed that far away from the nominal beam centre. Therefore, the beam was steered approximately 1$^{ \circ }$ away from its nominal position, and the TPC placed 1.5$^{ \circ }$ away from the nominal beam centre so that the TPC active region subtended an off-axis angular range of 1.4--3.8$^{ \circ }$.

There were four ToF constituents: 
\begin{itemize}
    \item $\mathit{S1}$, a small-area beam trigger, see Section~\ref{subsec:s1s2Exp};
    \item $\mathit{S2}$, a coincidence measurement with $\mathit{S1}$, see Section~\ref{subsec:s1s2Exp};
    \item $\mathit{S3}$, a panel of plastic scintillator bars placed directly upstream of the TPC vessel, see Section~\ref{subsec:s3Exp};
    \item $\mathit{S4}$, a panel of plastic scintillator bars placed directly downstream of the TPC vessel, see Section~\ref{subsec:s4Exp}.
\end{itemize}

\begin{figure}[H]
  \centering
  \includegraphics[width=1.0\linewidth]{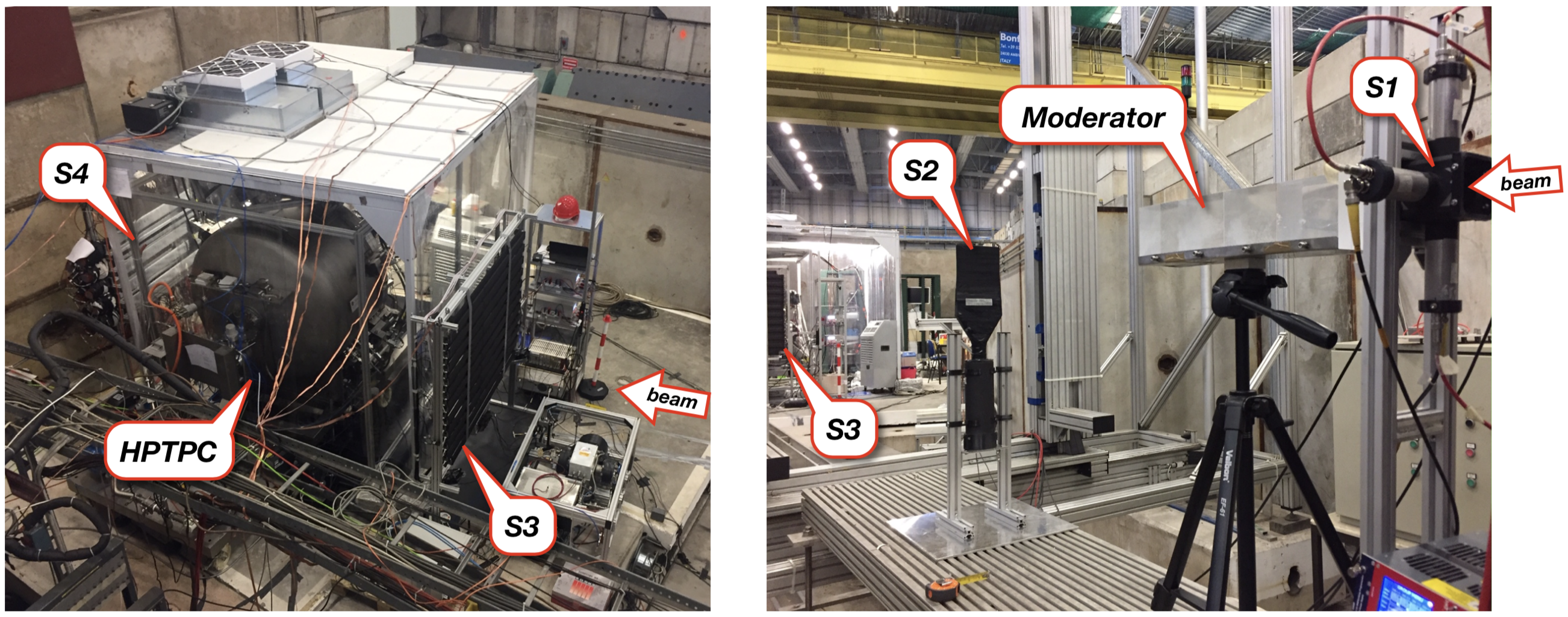}
  \caption{Photos illustrating the time of flight (ToF) constituents. (\textbf{Left}) the downstream part of the setup which shows the $\mathit{S3}$, $\mathit{S4}$ detectors and HPTPC. (\textbf{Right}) $\mathit{S1}$ and $\mathit{S2}$ counters and the stand with four acrylic moderator~blocks.}
  \label{fig:modblocks}
\end{figure}

A series of acrylic (polymethyl methacrylate) blocks was placed between the $\mathit{S1}$ and $\mathit{S2}$ counters.
Up to four $10\times10\times10$~cm$^3$ acrylic blocks could be placed contiguously on a tripod stand.
Figure~\ref{fig:modblocks} shows the stand with four blocks installed.
The moderator blocks have the effect of both reducing the energies of incoming particles as well as changing their directions.
This tends to increase the proton-to-MIP ratio at low off-axis angles from the beam, while decreasing the total number of protons and MIPs traversing the TPC.
Data were collected with the T10 beam momentum setting at 0.8~GeV/c, and with each configuration of 0 to 4 moderator blocks.

The data acquisition (DAQ) systems of the $\mathit{S3}$ (upstream) and $\mathit{S4}$ (downstream) ToF systems were completely independent.
Synchronization between ToF DAQ systems was performed offline using the reference signal from the PS at the beginning of every spill.
T10 received 1--3 spills from the PS during each supercycle, which has a typical duration of 33~s.
The spill duration is 400~ms.
The minimum separation in time between two spills is 1 s, so the start-of-spill signal frequency is less than or equal to 1~Hz.
As a result of the low frequency of start-of-spill signal, it is possible to use it, along with the DAQ file timestamps, to ensure that all spills are matched in both DAQs.
The trigger condition of the upstream ToF was based on the coincidence between $\mathit{S1}$ and $\mathit{S3}$ constituents.
$\mathit{S2}$ signals were also recorded by the upstream ToF DAQ but were not used in the trigger.
The DAQ of the downstream ToF was run in self-triggering mode with a gate open during the spill.
Coincidence signals between $\mathit{S1}$ and $\mathit{S2}$ counters were also recorded by the downstream ToF DAQ and were used in the particle identification (PID) analysis, described in Section~\ref{hptpcPaper:sec:Results}.  

\subsection{Survey and Coordinate System}
\label{sec:coord}
The T10 beamline area was surveyed, and the distances to specific components measured with \mbox{a precision} of 0.5~mm by the CERN Survey, Mechatronics and Measurements (SMM) group.
\mbox{Multiple points} on each of $\mathit{S1}$, $\mathit{S2}$, $\mathit{S3}$, $\mathit{S4}$ and the TPC frame have had their positions measured.

The axes of a right-handed coordinate system are defined as follows: $\hat{x}$ refers to the non-beam horizontal direction, $\hat{y}$ to the vertical direction, and $\hat{z}$ the beam direction, as shown in Figure~\ref{fig:setup}.
\mbox{We show} results in terms of two off-axis angles: $\theta$, which is measured in the $\hat{x}-\hat{z}$ plane with positive angles measured in the $+\hat{x}$ direction, and $\phi$, which is measured in the $\hat{y}-\hat{z}$ plane with positive angles measured in the $+\hat{y}$ direction.
The origin is taken to be at $\mathit{S1}$.

\begin{figure}[H]
  \begin{adjustbox}{max totalsize={0.7\textwidth}{.5\textheight},center}
    \input{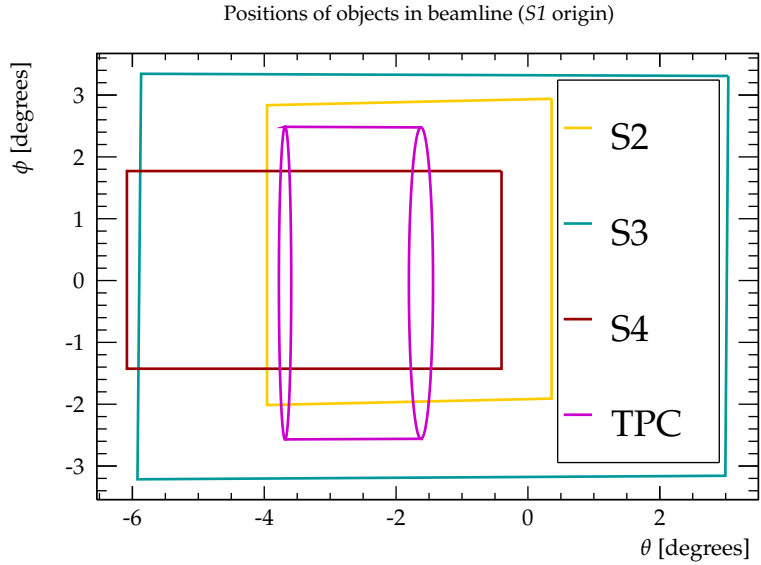}
  \end{adjustbox}
  \caption{Angular position of various objects within the T10 beamline. The origin in this view is at the centre of $\mathit{S1}$; the true centre of the steered beam is at +1$^{\circ}$ in $\theta$ and 0$^{\circ}$ in $\phi$.}
  \label{fig:angularDistS1}
\end{figure}

Figure~\ref{fig:angularDistS1} shows the angular extent of objects within the beamline using the coordinate system defined above.
Table~\ref{tab:angS1} shows the calculated angular extent of the various beamline components as measured from $\mathit{S1}$.
Table~\ref{tab:distances} shows the distances between the centres of various objects in the T10 beamline.
These distances were calculated using the data gathered by the survey team.

\begin{table}[H]
  \centering
  \caption{Angular extents of objects within the T10 beamline as measured from $\mathit{S1}$.}
  \begin{tabular}{ccccc}
    \toprule
    \textbf{Object} & \textbf{Minimum \boldmath{$\theta$}} & \textbf{Maximum \boldmath{$\theta$}} & \textbf{Minimum \boldmath{$\phi$}} & \textbf{Maximum \boldmath{$\phi$}} \\
    \midrule
    $\mathit{S2}$ & $-3.96^{\circ} \pm 0.03^{\circ}$ & $0.36^{\circ} \pm 0.03^{\circ}$ & $-2.01^{\circ} \pm 0.03^{\circ}$ & $2.94^{\circ} \pm 0.03^{\circ}$ \\
    $\mathit{S3}$ & $-5.923^{\circ} \pm 0.004^{\circ}$ & \hspace{6pt} $3.040^{\circ} \pm 0.004^{\circ}$ & $-3.215^{\circ} \pm 0.004^{\circ}$ & $3.344^{\circ} \pm 0.004^{\circ}$ \\
   $\mathit{S4}$ & $-6.083^{\circ} \pm 0.003^{\circ}$ & $-0.401^{\circ} \pm 0.003^{\circ}$ & $-1.426^{\circ} \pm 0.003^{\circ}$ & $1.771^{\circ} \pm 0.003^{\circ}$ \\
    TPC upstream face & $-3.59^{\circ} \pm 0.01^{\circ}$ & $-1.44^{\circ} \pm 0.01^{\circ}$ & $-2.66^{\circ} \pm 0.01^{\circ}$ & $2.58^{\circ} \pm 0.01^{\circ}$ \\
    TPC downstream face & $-3.778^{\circ} \pm 0.009^{\circ}$ & $-1.806^{\circ} \pm 0.009^{\circ}$ & $-2.440^{\circ} \pm 0.009^{\circ}$ & $2.361^{\circ} \pm 0.009^{\circ}$ \\
    \bottomrule
  \end{tabular}
  \label{tab:angS1}
\end{table}

\begin{table}[H]
  \centering
  \caption{Distances between objects in the T10 beamline. US and DS refer to the upstream and downstream edges of the TPC, respectively.}
 \begin{tabular}{cc}
    \toprule
    \textbf{Points} & \textbf{Distance between Centres/m}\\
    \midrule
    Beam monitor -- $\mathit{S1}$ & $0.288 \pm 0.001$ \\
    $\mathit{S1}-\mathit{S2}$ & $1.419 \pm 0.001$ \\
    $\mathit{S1}-\mathit{S3}$ & $10.756 \pm 0.001$ \\
    $\mathit{S3}$ -- TPC US side & $1.323 \pm 0.002$ \\
    TPC DS side -- $\mathit{S4}$ & $0.918 \pm 0.002$ \\
    $\mathit{S2}-\mathit{S4}$ & $12.651 \pm 0.001$ \\
    \bottomrule    
  \end{tabular}
  \label{tab:distances}
\end{table}

\subsection{Upstream Beam Counters (S1 and S2)}
\label{subsec:s1s2Exp}

\begin{figure}[H]
  \centering
  \includegraphics[width=0.65\linewidth]{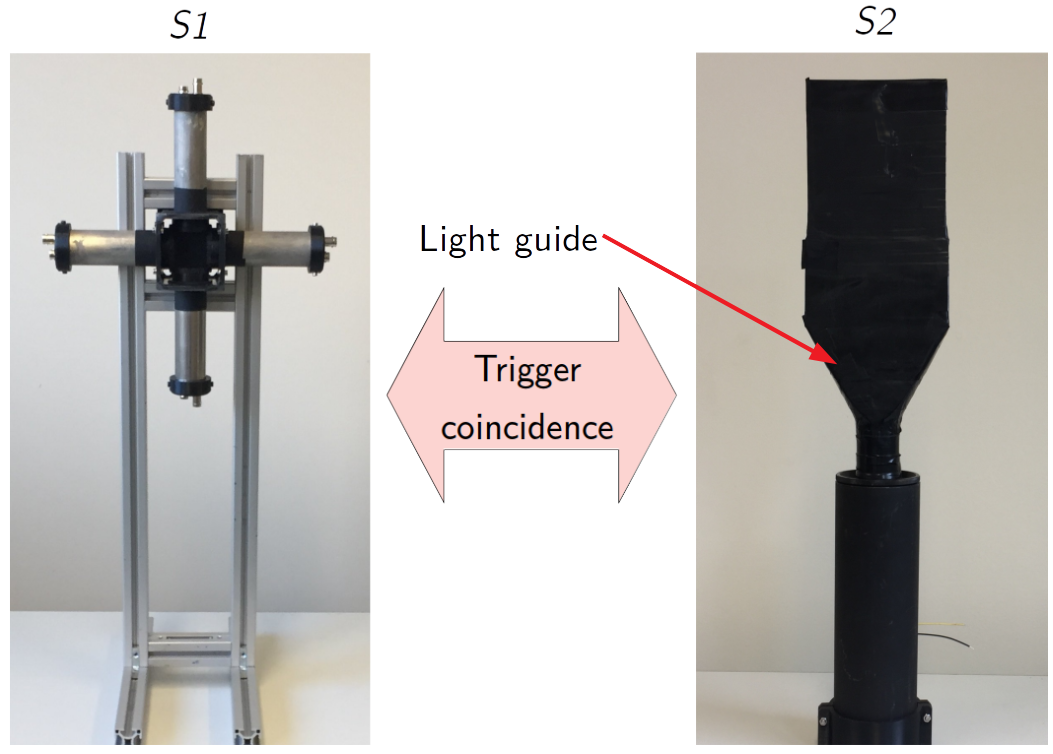}
  \caption{The S1 and S2 beam counters. Together the coincidence of signals in the beam counters were recorded by the data acquisition (DAQ) systems.}
  \label{fig:S1S2headon}
\end{figure}

\begin{figure}[H]
  \begin{adjustbox}{width=0.65\linewidth, center}
    \input{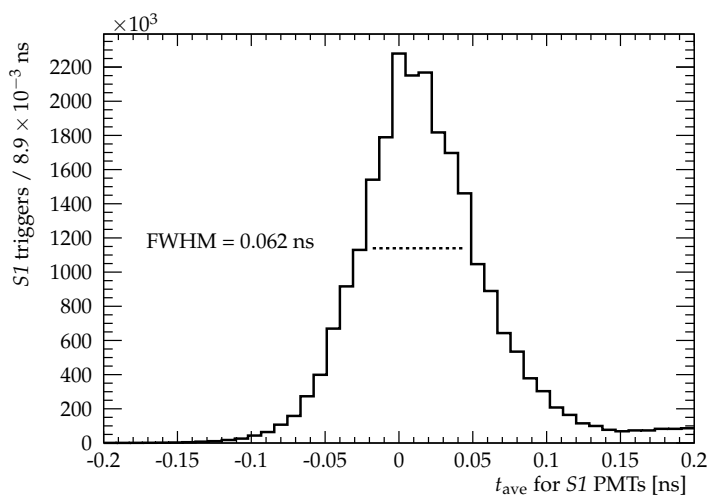}
  \end{adjustbox}
  \caption{Example of the timing spread of $\mathit{S1}$ hits. The time is calculated as an average of the hit time as measured in each of the four photomultiplier tubes (PMTs).}
  \label{fig:s1Res}
\end{figure}

The beam counters $\mathit{S1}$ and $\mathit{S2}$ are shown in Figure~\ref{fig:S1S2headon}.
The $\mathit{S1}$ counter is a $40\times40\times5$~mm$^3$ plastic scintillator cross which is attached to four 1'' Hamamatsu Photonics R4998 photomultiplier tubes (PMTs) at each end for the light readout.
The time resolution of the counter, as measured by the DAQ system of the upstream ToF, was about 30~ps. This is estimated with the distribution of the average PMT hit times; the quantity $t_{\textrm{ave}}=\frac{1}{4}((t_{\textrm{PMT0}}+t_{\textrm{PMT1}})-(t_{\textrm{PMT2}}+t_{\textrm{PMT3}}))$ has the same spread as the simple average but is conveniently centred at zero.
An example of the $t_{\textrm{ave}}$ distribution for one run of $\mathit{S1}$ data is shown in Figure~\ref{fig:s1Res}. The full width at half maximum (FWHM) of the distribution is 62 ps.

\unskip

The $\mathit{S2}$ counter is a scintillator tile of size $120\times120\times5$~mm$^3$, coupled to a 2'' Hamamatsu Photonics R1309 PMT~\cite{Hamamatsu}, via a long light-guide as shown in Figure~\ref{fig:modblocks}.
The $\mathit{S2}$ counter was placed $(1.419 \pm 0.001)~\text{m}$ downstream of $\mathit{S1}$.
The transverse position of $\mathit{S2}$ was adjusted to account for the beam divergence in the moderator blocks.

The analog signals from one of the $\mathit{S1}$ PMTs and the $\mathit{S2}$ PMT were fed into LeCroy 620AL NIM discriminator units with a threshold of 30~mV.
Subsequently, the discriminated signals were fed into a NIM coincidence unit, whose output was recorded by the DAQ systems of the downstream ToF ($\mathit{S4}$) panel.
This information was further used for the time of flight analysis of $\mathit{S4}$.

\subsection{Upstream Time of Flight Instrumentation (S3)}
\label{subsec:s3Exp}
\begin{figure}[H]
  \centering
  \includegraphics[width=0.52\linewidth]{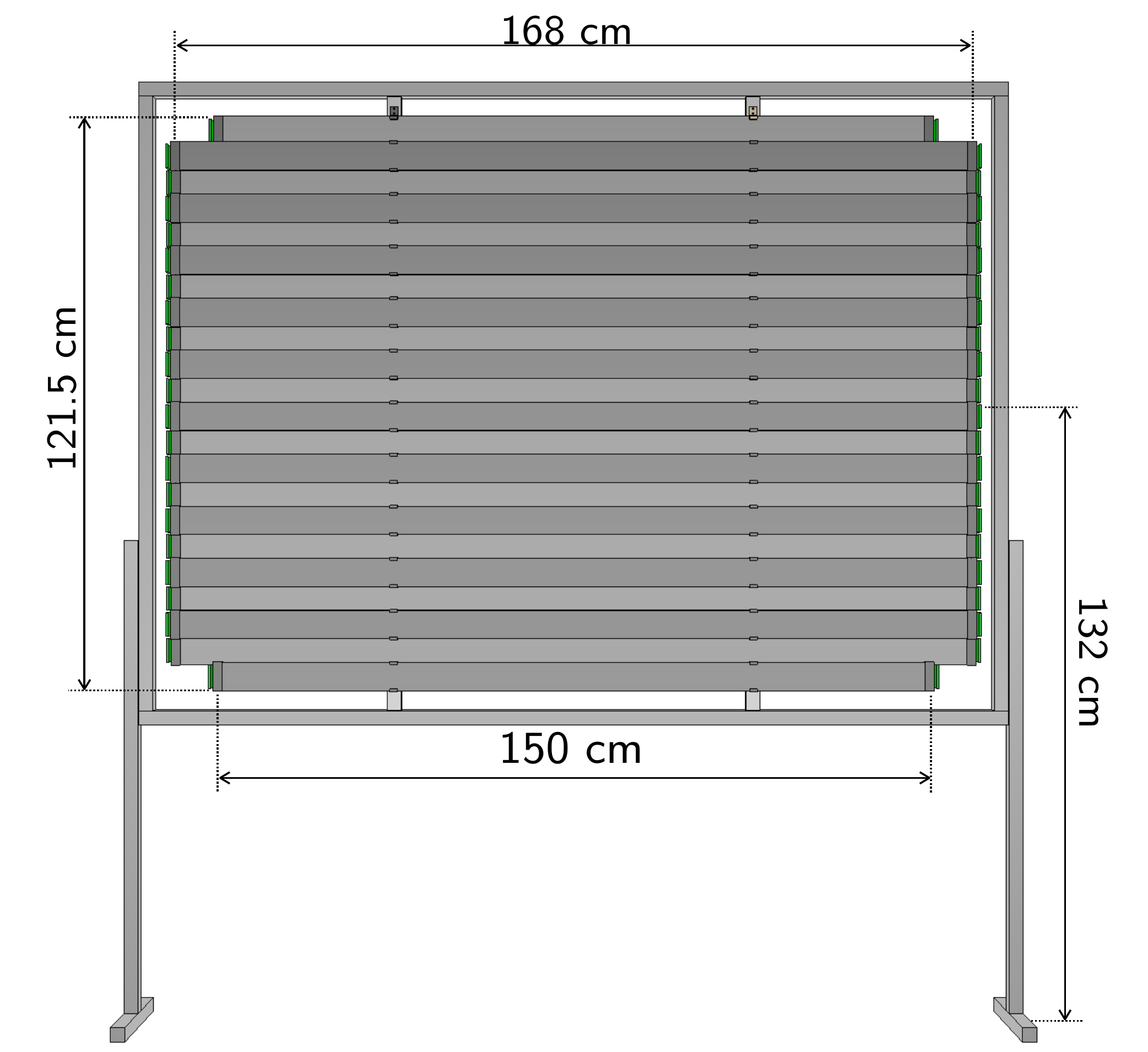}
  \hfill
  \includegraphics[width=0.47\linewidth]{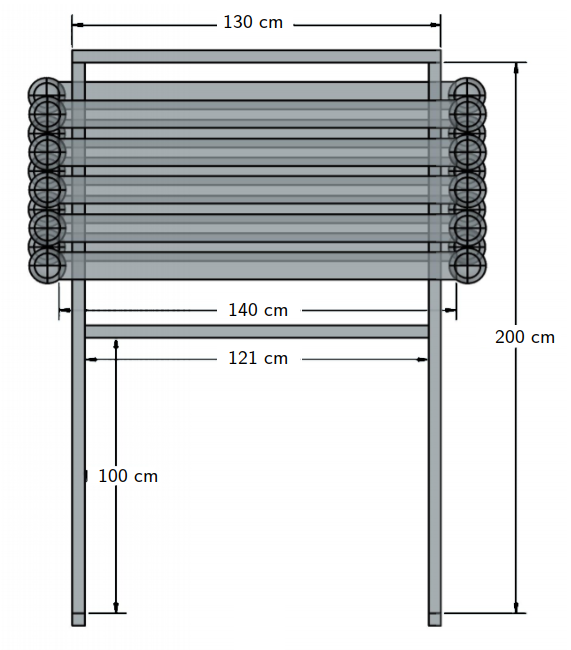}
  \caption{View of the time of flight panels.
  (\textbf{Left}) The $\mathit{S3}$ panel~\cite{S3-proceedings} upstream of the TPC. (\textbf{Right}) The $\mathit{S4}$ panel downstream of the TPC.}
  \label{fig:S3sketch}
\end{figure}

The $\mathit{S3}$ `upstream' ToF constituent was placed $(1.323 \pm 0.001)~\text{m}$ upstream of the upstream side of the HPTPC drift volume in the beamline.
A schematic drawing of the $\mathit{S3}$ ToF panel is shown in Figure~\ref{fig:S3sketch} left.
The detector comprises 22 staggered scintillator bars:  20 bars with dimensions $168 \times 6.0 \times 1.0$~cm$^3$ and 2 bars of  $150 \times 6.0 \times 1.0$~cm$^3$ placed on top and bottom~\cite{S3-proceedings}.
The overlap between bars was set to 5~mm, thus the active area of the detector was $2.0214~\text{cm}^{2}$.

The bars are made from EJ-200~\cite{SCIONIX} plastic scintillator, which provides a brightness of 10,000~photons/MeV~deposited.
It also has a suitable optical attenuation length of 4~m and fast timing, with a rise time of 0.9~ns and decay time constant of 2.1~ns.
The scintillation emission spectrum of EJ-200 peaks in the violet region of the visible spectrum (435~nm)~\cite{EJ200}.
The bars were wrapped in an aluminium foil (60\% reflectivity) to increase the collected light.

Arrays of eight $6 \times 6$~mm$^2$ area silicon photomultipliers (SiPMs) S13360-6050PE from Hamamatsu Photonics \cite{Hamamatsu} were coupled to each end of the bar to collect scintillation photons.
The photon detection efficiency at the peak sensitivity wavelength (450~nm) is 40\%~\cite{Hamamatsu}.
The anode signals of the SiPMs are read out, summed and shaped by a dedicated circuit as described in Ref.\,\cite{S3-readout}.

$\mathit{S3}$ uses a 64 channel data acquisition system based on the SAMPIC chip.
A SAMPIC chip is \mbox{a waveform} and time to digital converter (WTDC) 16-channel ASIC which provides a raw time with ultrafast analog memory allowing fine timing extraction as well as other parameters of the pulse~\cite{SAMPIC}.
Each channel contains a discriminator that can trigger itself independently or participate in a more complex combined trigger. 
Three ASIC modules ($16\times3=48$ channels) were connected to the \mbox{44 channels} of $\mathit{S3}$ and were operated in self-triggering mode.

The trigger conditions are as follows: at least three out of the four $\mathit{S1}$ PMTs must have a signal above a 30~mV threshold.
Additionally, there must be at least one signal in $\mathit{S3}$ above 30~mV.
These~$\mathit{S1}$ and $\mathit{S3}$ signals must be coincident within a gate of 70~ns.
A fourth ASIC was used to acquire data from $\mathit{S1}$, the coincidence signal $\mathit{S1} \cap \mathit{S2}$ and the start-of-spill signal from the PS.
The mean time of light signals detected at both ends of a single bar provides a time reference with a resolution of about 100~ps, while the difference between the time of the light signals gives the position of the interaction along the bar, with a resolution of 1.6~cm.

\begin{figure}[H]
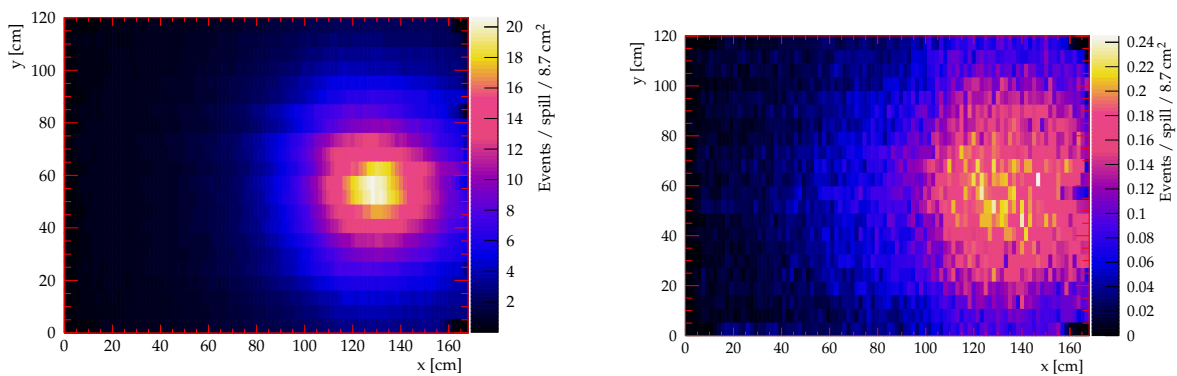

  \begin{minipage}[t]{0.49\textwidth}
    \centering
    \begin{adjustbox}{max totalsize={\textwidth}{.5\textheight},center}
      \input{files/Figures/4_pionXY.tex}
    \end{adjustbox}
  \end{minipage} 	
  \hfill
  \begin{minipage}[t]{0.49\textwidth}
    \centering
    \begin{adjustbox}{max totalsize={\textwidth}{.5\textheight},center}
      \input{files/Figures/4_protonXY.tex}
    \end{adjustbox}
  \end{minipage}  
   \caption{ \label{fig:s3XY_pion}Reconstructed positions of hits observed in $\mathit{S3}$. (\textbf{Left}) Minimum ionizing particles with four moderator blocks placed in the beamline. (\textbf{Right}) Protons detected with four moderator blocks placed in the beamline. This figure uses local $\mathit{S3}$ coordinates in which $y,x=0~\text{cm}$ is the bottom right corner of the active area when viewed from $\mathit{S1}$.}
\end{figure}

Examples of reconstructed $\mathit{S3}$ spatial distributions are shown in Figure~\ref{fig:s3XY_pion}.
Figure~\ref{fig:s3XY_pion} left, shows the spatial distribution of hits in $\mathit{S3}$ thought to be produced by MIPs when 4 moderator blocks were in the beamline.
Figure~\ref{fig:s3XY_pion} right, shows the spatial distribution of hits identified in $\mathit{S3}$ as protons when 4 moderator blocks were in the beamline.
The pattern of hits is more diffuse, illustrating the scattering effect of the moderator blocks.
When in this position, the measured horizontal FWHM of the unmoderated beam is 16.8~cm while the vertical FWHM is 11.0~cm.
With 4 moderator blocks in the beamline, the measured horizontal FWHM of the beam is 63.8~cm while the vertical FWHM is 60.0~cm.

\begin{figure}[H]
  \centering
  \includegraphics[width=.8\linewidth]{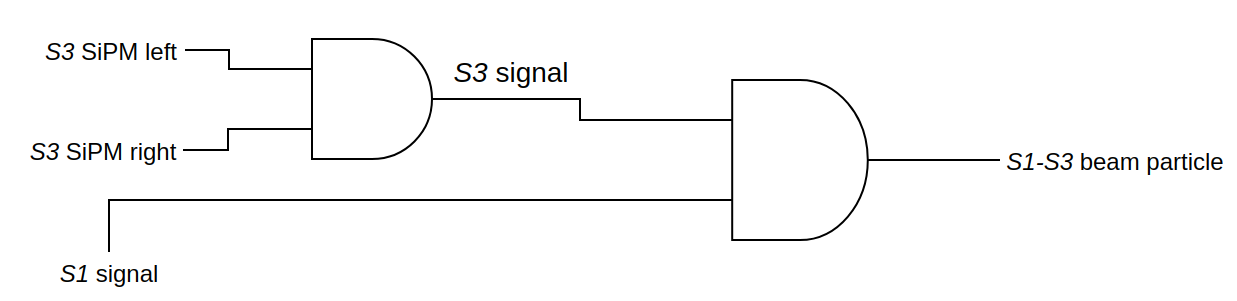}
  \caption{Simplified trigger logic diagram for the upstream ToF detection of a beam particle, showing the required coincidences. Left and right refer to the silicon photomultipliers (SiPMs) on the opposite ends of the same bar.}
  \label{fig:utofTrig}
\end{figure}

Figure~\ref{fig:utofTrig} shows the required trigger logic for the detection of a beam particle in the upstream ToF instrumentation.
The signal thresholds and timing cuts used for the coincidences are those detailed in this section.

\subsection{Downstream Time of Flight Instrumentation (S4)}
\label{subsec:s4Exp}
The $\mathit{S4}$ `downstream' ToF constituent sat $(0.918 \pm 0.001)~\text{m}$ downstream of the downstream edge of the drift volume of the HPTPC prototype in the beamline.
It consists of 10 bars of Nuvia NuDET plastic scintillator which has a wavelength of maximum emission of 425~nm and a decay time constant of 2.5~ns~\cite{Nuvia}.
Each of these bars measure $10 \times 1 \times 140$~cm$^3$. 
Attached to each end of these scintillator bars is a 5" Hamamatsu Photonics R6594 PMT~\cite{Hamamatsu}.
The bars are arranged in two rows of five, such that there is complete coverage for any beam particles incident upon the detector.
The bars are wrapped individually in reflective milar sheets to increase the light yield.
The total active area of the $\mathit{S4}$ panel is $1.40 \times 0.78$~m$^2$.
A diagram of $\mathit{S4}$ along with its dimensions is presented in Figure~\ref{fig:S3sketch} right.

\begin{figure}[H]
  \begin{adjustbox}{max totalsize={.6\textwidth}{.5\textheight},center}
    \input{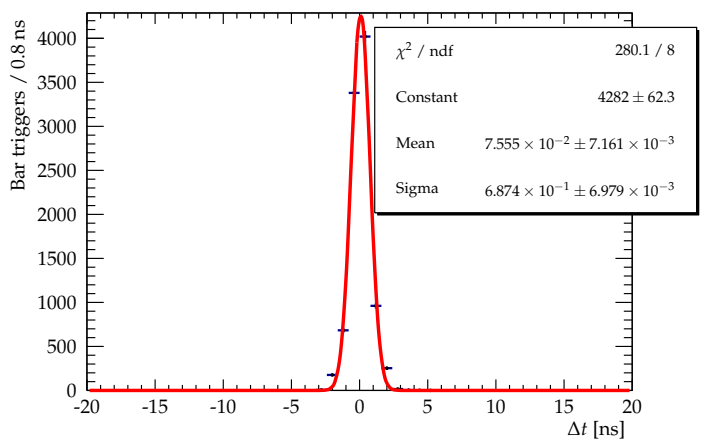}
  \end{adjustbox}
  \caption{Difference in signal arrival time for PMTs at each end of a bar as measured using a $^{90}$Sr source placed 64~cm from one end of the bar.}
  \label{fig:s4Res}	
\end{figure}

The time resolution of the bars and PMTs is measured to be 0.8~ns using a $^{90}$Sr source placed at measured distances along the bar.
Figure~\ref{fig:s4Res} is the measured time difference for signals coming from the PMTs at either end of a bar caused by the $^{90}$Sr at a given position.
Figure~\ref{fig:s4Res} shows an example of the distribution from which the time resolution was derived.
The corresponding spatial resolution of the bars and PMTs was measured to be 7~cm.

The anode signals of all 20 of the PMTs are discriminated using LeCroy 620AL NIM discriminators, at a threshold of 20~mV.
The discriminated signals are then fed into a time-to-digital converter (TDC). A signal in $\mathit{S4}$ is deemed to have occurred if a signal is seen in both PMTs, above the discriminator threshold, on the same bar within 20~ns of each other. 
This timing window is determined through testing performed with a $^{90}$Sr source at known positions on the bar.

The $\mathit{S1-S2}$ coincidence signal is digitized by the same TDC. This signal is used to calculate the particle time of flight from $\mathit{S2}$ to $\mathit{S4}$.

\subsection{The HPTPC Prototype}
For the characterisation of the beam using the ToF systems described in this paper, the relevant characteristics of the HPTPC prototype are the location and thickness of the steel vessel walls.
\mbox{The cylindrical} steel vessel has a 142~cm outer diameter; the main body is 60 cm in length and the rounded end caps protrude an additional 37~cm on each end.
With 1~cm thick walls it is rated to 6~bar of absolute pressure.
The vessel wall thickness is equivalent to the range of a proton with a kinetic energy of approximately 80~MeV~\cite{rangeTables}.
For the unmoderated beam, the typical energy loss of a proton which does not stop in the vessel is 50~MeV.
This is determined from the Monte Carlo studies detailed in Section~\ref{sec:mcStudies}.
The angular position of the centre of the TPC is approximately $\theta = -2.5^{\circ}$. 
More details of the position and extent of the TPC are given in Tables~\ref{tab:angS1} and~\ref{tab:distances}.
 
\begin{figure}[H]
  \centering
  \includegraphics[width=.8\linewidth]{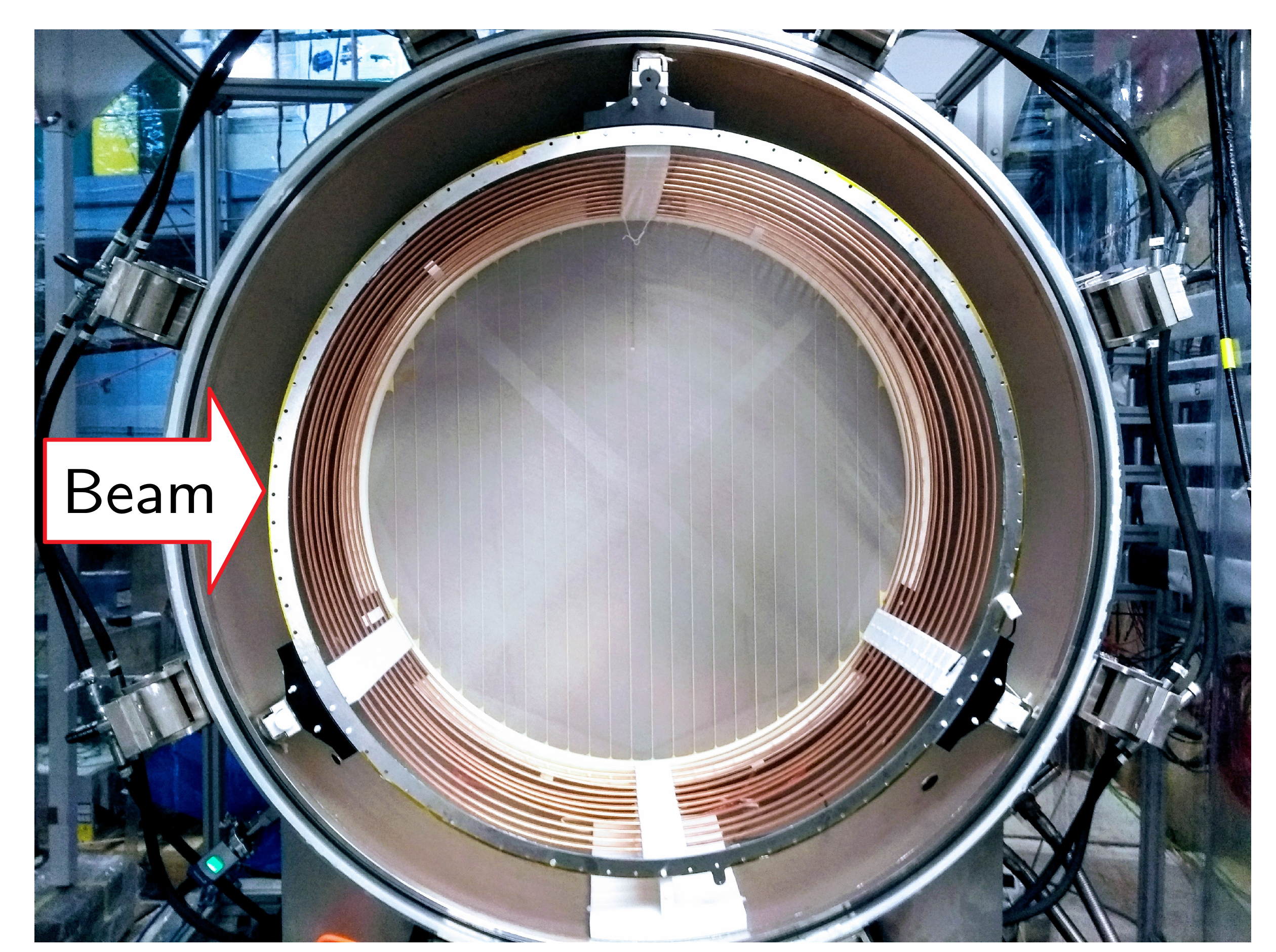}
  \caption{Cross-sectional view of the TPC; the thin mesh electrodes and copper ring drift volume can be seen inside the steel vessel. The walls of the vessel shown are 1~cm thick with a vessel outer diameter of 142~cm. At the point of hitting the vessel, the beam centre was 1~cm below the centre of the vessel vertically, where the distance from the inside of the vessel wall to the drift region was 15~cm.}
   \label{fig:TPC}
\end{figure}

The active TPC is a cylinder, 111~cm in diameter and 48~cm in length; the TPC comprised thin steel mesh electrodes (one cathode with \SI{118}{\centi\meter} diameter and three anodes with \SI{121}{\centi\meter} diameter), \mbox{and 12 copper} rings to create the uniform drift field. The anodes were supported by a hexagonal aluminium stiffener on the side facing away from the camera.
Data taking with the TPC made use of both optical and charge readout.
The vessel, electrodes and drift region of the TPC are shown in Figure~\ref{fig:TPC}.

Throughout the run, the TPC was filled with either pure argon, or a combination of argon and a small percentage of quencher.
The performance of this TPC is the subject of \mbox{a forthcoming~publication~\cite{Deisting:2020aaa}}.

\section{Analysis}\vspace{-6pt}
\label{hptpcPaper:sec:Analysis}
\subsection{Analysis Goals}

The primary aims of this analysis are to assess the feasibility of using the combination of off-axis positioning and a moderated beam to produce particles with momenta covering the range of momenta of particles produced in GeV-scale neutrino interactions and to characterize the incident flux on the TPC and exiting the TPC, for the TPC data analysis.

The numbers of spills recorded for each number of moderator blocks are shown in Table~\ref{tab:spills}.
More~data were collected for 4 blocks as that was the configuration used for the majority of the beam~test.

\begin{table}[H]
  \centering
  \caption{Total number of spills recorded for each moderator block configuration included in this paper.}
  \begin{tabular}{cc}
    \toprule
    \textbf{Number of Moderator Blocks} & \textbf{Recorded Spills} \\
    \midrule
    0 & 257 \\
    1 & 254 \\
    2 & 267 \\
    3 & 220 \\
    4 & 3884 \\
    \bottomrule
  \end{tabular}
  \label{tab:spills}
\end{table}

\subsection{Time of Flight Analysis}
\label{timeofflightanalysissubsec}

\begin{figure}[H]
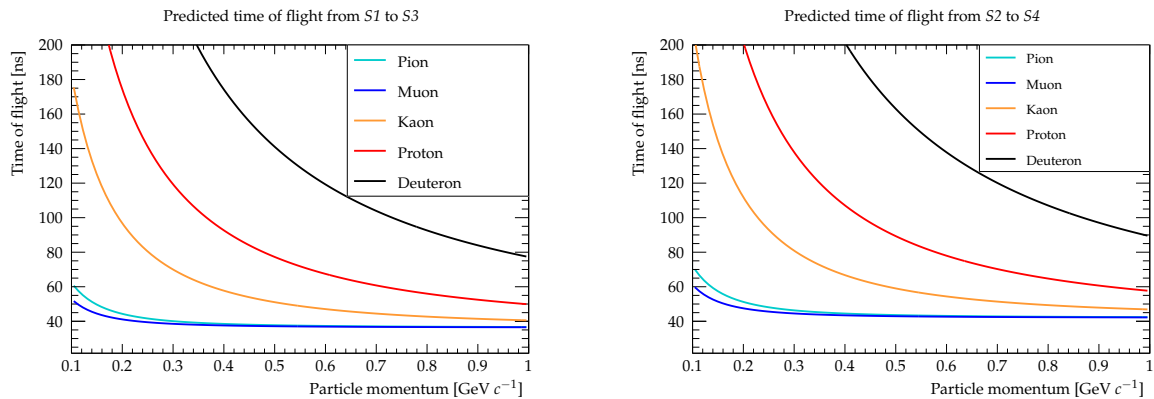

  
  \begin{minipage}[t]{0.49\textwidth}
    \begin{adjustbox}{max totalsize={\textwidth}{.5\textheight},center}
      \input{files/Figures/s1s3Times.tex}
    \end{adjustbox}
  \end{minipage}
  \hfill
  \begin{minipage}[t]{0.49\textwidth}
    \begin{adjustbox}{max totalsize={\textwidth}{.5\textheight},center}
      \input{files/Figures/s2s4Times.tex}
    \end{adjustbox}
  \end{minipage}
   \caption{\label{fig:s1s3PredTimes}Calculated time of flight for a number of different particle species as a function of particle momentum. (\textbf{Left}) ToF between $\mathit{S1}$ and $\mathit{S3}$. (\textbf{Right}) ToF between $\mathit{S2}$ and $\mathit{S4}$.}
\end{figure}

A charged pion with a momentum of 0.8~GeV/c will have a time of flight from $\mathit{S1}$ to $\mathit{S3}$ (a distance of 10.8~m) of 37~ns, while a proton with the same momentum will have a time of flight of 55~ns.
\mbox{For the} same two particles travelling between $\mathit{S2}$ and $\mathit{S4}$ (a distance of 12.7~m), the charged pion would have a time of flight of 43~ns and the proton would have a time of flight of 65~ns.
Figure~\ref{fig:s1s3PredTimes} left and right, shows the predicted time of flight for various particle species across the $\mathit{S1}-\mathit{S3}$ distance and the $\mathit{S2}-\mathit{S4}$ distance, respectively.

\begin{figure}[H]
  \begin{adjustbox}{max totalsize={.7\textwidth}{.5\textheight},center}
    \input{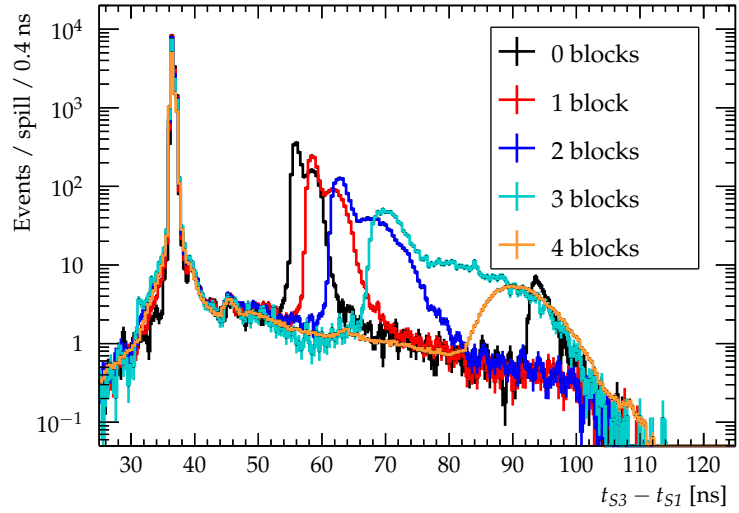}
  \end{adjustbox}
  \caption{$\mathit{S3}$ time of flight spectra for varying numbers of moderator blocks.}
  \label{fig:s3tof}
\end{figure}

Figure~\ref{fig:s3tof} shows the time of flight spectrum recorded in the $\mathit{S3}$ timing point for varying numbers of moderator blocks.
The quicker peak is formed by minimum ionizing particles, while the peak at higher values of $\mathit{t_{S3}} - \mathit{t_{S1}}$ corresponds to protons.
The proton peaks show a double peak feature, with a smaller delayed peak closely following the main proton peak; this feature appeared after the beam was steered so that the full 2.5$^{ \circ }$ off-axis angle could be achieved and is due to a portion of beam scattering in the steering magnets, leading to the slower peak.
The part of the beam which does not impinge on the steering magnets produces the quicker proton peak in the spectrum.
Figure~\ref{fig:utofNoBend} left and right, shows the proton peak for unsteered beam and the double peak structure is gone.
In the black curve, which shows the 0 block data, a deuteron peak can be seen centred at 95~ns.
The timing ranges for particle species selection are chosen using the analytic expectations shown in Figure~\ref{fig:s1s3PredTimes}.

To calculate the correct time of flight, timing delays caused by cabling and equipment are taken into account.
The same method is used to correct the measurements of the time of flight between $\mathit{S1}$ and $\mathit{S3}$, and $\mathit{S2}$ and $\mathit{S4}$.
The initial recorded time, $t_i$, is either $t_{\mathit{S1}}$ or $t_{\mathit{S2}}$ while the final recorded time, $t_f$ is then $t_{\mathit{S3}}$ or $t_{\mathit{S4}}$, respectively.

Timing offsets are measured in the beamline by assuming that the fastest peak in the $t_{f}-t_{i}$ spectrum for the unmoderated data is produced by charged MIPs with a momentum of 0.8~GeV/c.
\mbox{The required} timing shift is then the shift required to move the fastest peak to its expected position, \mbox{given this} assumption.
This shift is then applied to all measured times of flight.
This correction is peformed separately for both the measurement of $t_{\mathit{S3}}-t_{\mathit{S1}}$ and for the measurement of $t_{\mathit{S4}}-t_{\mathit{S2}}$.
The~required timing shift for the $t_{\mathit{S4}}-t_{\mathit{S2}}$ measurement is 43.7~ns.
For the $t_{\mathit{S3}}-t_{\mathit{S1}}$ measurement, the required timing shift is 65.0~ns.

\begin{figure}[H]
\centering
  \begin{adjustbox}{max totalsize={\linewidth}, center}
    \input{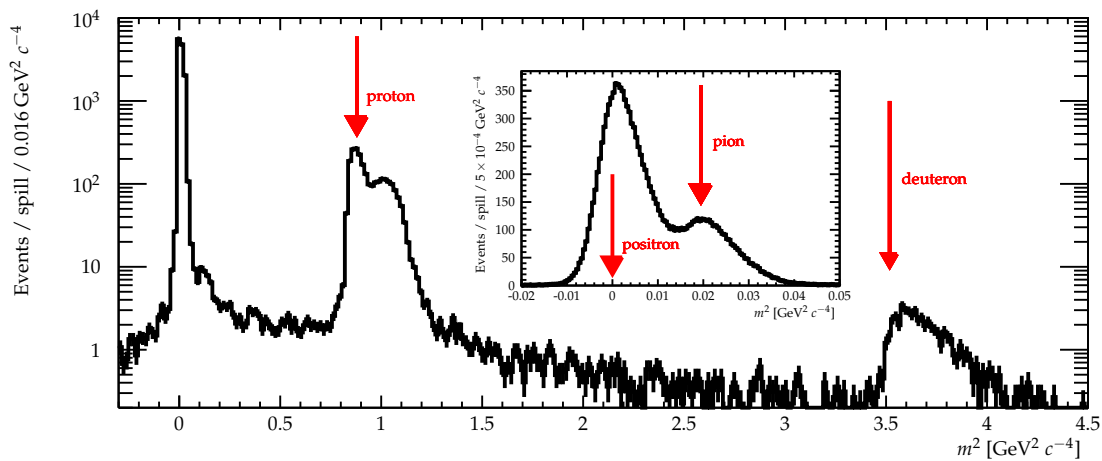}
  \end{adjustbox}
  \caption{Reconstructed mass spectrum for the data taken without moderator blocks. The spectrum was calculated using the time difference between $\mathit{S3}$ and $\mathit{S1}$. Vertical arrows show predicted position of particles given a momentum of 0.8~GeV/c. Insert: Zoomed view of MIP region of the same spectrum.}
  \label{fig:s3tof_mass}
\end{figure}

The mass distribution calculated for the dataset without moderator blocks is shown in Figure~\ref{fig:s3tof_mass}.
The time difference between $\mathit{S3}$ and $\mathit{S1}$ counters corresponding to a single particle ($t_{\mathit{S3}}-t_{\mathit{S1}}$) is converted to the mass of the particle, $m$, using Equation~\eqref{eq:recoMass}, where the equation is in natural units.
\mbox{The particle} momentum, $p$, is assumed to be 0.8~GeV/c.

\begin{equation} 
  m^2 = p^2 \left( 
  \left(\frac{t_{\mathit{S3}}-t_{\mathit{S1}}}{x_{\mathit{S3}}-x_{\mathit{S1}}} \right)^2 - 1  \right) \,,
  \label{eq:recoMass}
\end{equation}

The proton and pion mass positions in Figure~\ref{fig:s3tof_mass} are indicated by vertical arrows.
One can clearly observe distinct peaks corresponding to protons and deuterons. 
The insert in the figure shows \mbox{a zoomed} region corresponding to the MIPs. 

\begin{figure}[h]
  \centering
  \includegraphics[width=.9\linewidth]{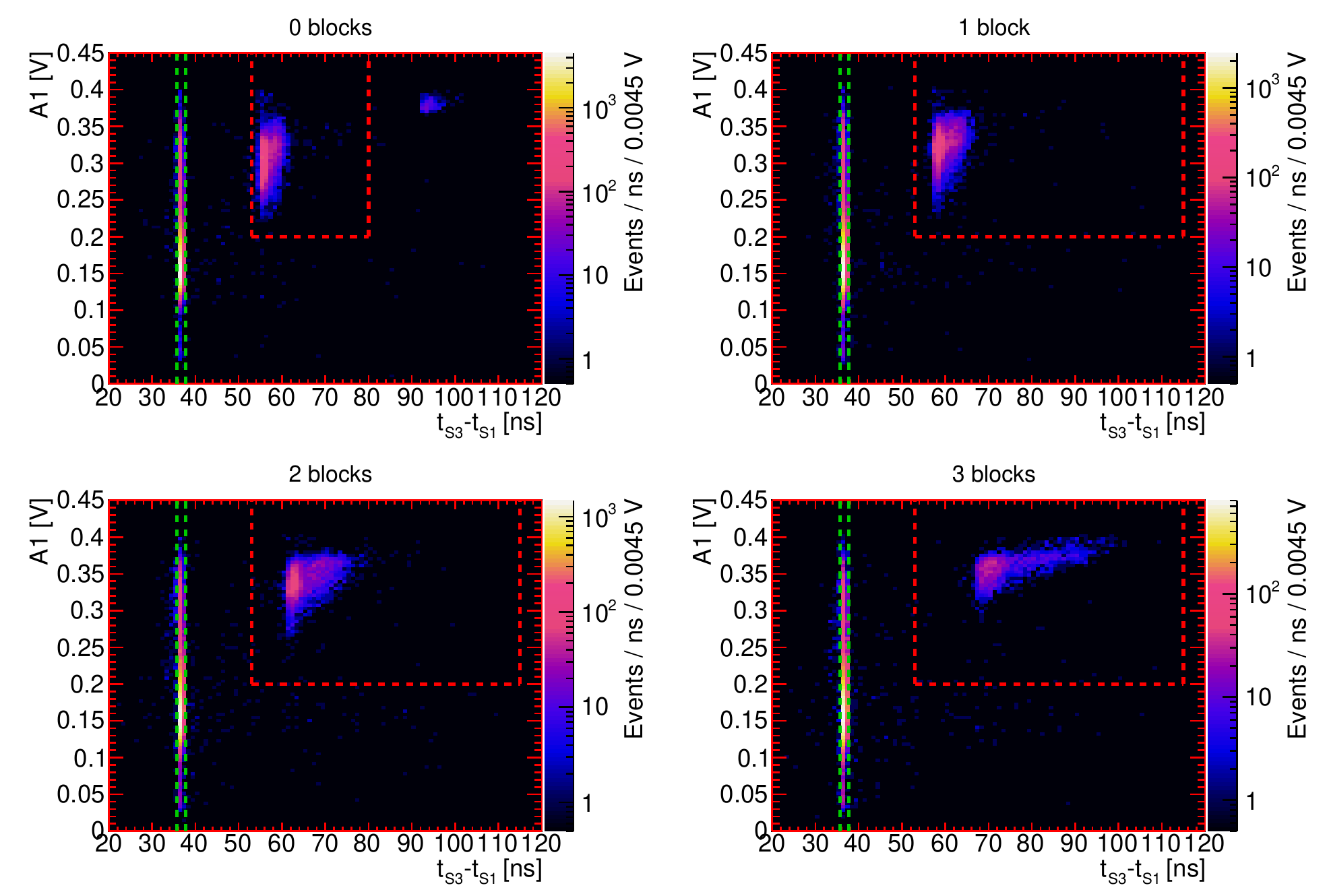}
  \caption{Examples of SiPM signal amplitude plotted against $\mathit{S1}$ to $\mathit{S3}$ time of flight for different numbers of moderator blocks. Clockwise from top left 0, 1, 3 and 2 moderator blocks are shown. \mbox{A1 is} the voltage recorded in the SiPM at the end of the bar. The red horizontal dashed line shows the amplitude cut used for this particular SiPM. Events in the area enclosed by the red dashed lines are selected as protons. Events enclosed by the green dashed lines are selected as MIPs.}
  \label{fig:TvsA}
\end{figure}

For the data collected in $\mathit{S3}$, both timing and signal amplitude cuts were used to select protons and MIPs.
Figure~\ref{fig:TvsA} shows an example of the signal size recorded in one of the SiPMs on one of the scintillator bars against the measured value of $t_{\mathit{S3}} - t_{\mathit{S1}}$.
At the beam energies used, due to their higher mass, the protons typically deposit more energy in the detector, resulting in the observation of greater amplitudes.
Therefore, to reduce the number of background events in the proton sample, a minimum signal amplitude is required.
This cut varies, depending on the SiPM in question and is determined from distributions such as those shown in Figure~\ref{fig:TvsA}. 
The cut values vary in the range 0.125~V to 0.3~V.

Particles for which $35.75~\text{ns}<t_{\mathit{S3}}-t_{\mathit{S1}}<37.75~\text{ns}$ are identified as MIPs.
Particles which pass the amplitude cut and for which $53~\text{ns}<t_{\mathit{S3}}-t_{\mathit{S1}}<115~\text{ns}$ are identified as protons.
The upper bound of this timing cut is reduced to 80~ns for the unmoderated sample in order to exclude deuterons.

A correction must be applied to the upstream ToF DAQ ($\mathit{S1}$, $\mathit{S2}$ and $\mathit{S3}$) to account for its large dead time.
The $\mathit{S1} \cap \mathit{S2}$ signal is digitised by both the upstream and downstream ToF DAQ.
The dead time of the downstream ToF DAQ is found to be negligible.
A linear relationship between the number of $\mathit{S1} \cap \mathit{S2}$ signals measured in each DAQ is determined for each moderator block sample.
\mbox{Therefore, events measured} in the upstream ToF DAQ are weighted, such that the number of $\mathit{S1} \cap \mathit{S2}$ signals measured in the upstream and downstream ToF DAQs are approximately equal.

\begin{figure}[H]
\centering
  \begin{adjustbox}{max totalsize={.83\textwidth}{.5\textheight},center}
    \input{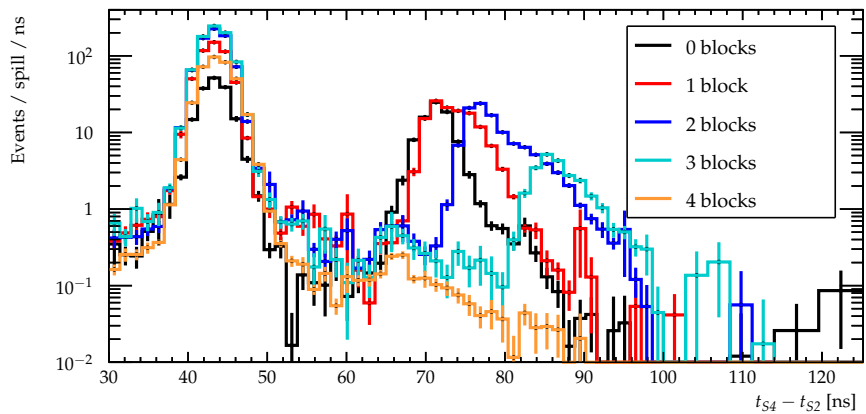}
  \end{adjustbox}
  \caption{$\mathit{S4}$ time of flight spectra for varying numbers of moderator blocks. For all configurations, a flat background has been fitted and subtracted from the data. Additionally, the plot has also been corrected for the differing efficiencies of the various bars and for the variation in efficiency as a function of position along the bar, as described in Section~\ref{timeofflightanalysissubsec}.}
  \label{fig:s4tof}	
\end{figure}

Figure~\ref{fig:s4tof} shows the variation in the time of flight spectrum as recorded by $\mathit{S4}$ with a changing number of moderator blocks.
This spectrum is given by the difference in time between observation of a coincidence in the $\mathit{S1}$ and $\mathit{S2}$ timing points and a signal being recorded in $\mathit{S4}$ (the definition of an $\mathit{S4}$ signal is given above).

\begin{figure}[H]
  \centering
  \begin{adjustbox}{max totalsize={.83\textwidth}{.5\textheight},center}
    \input{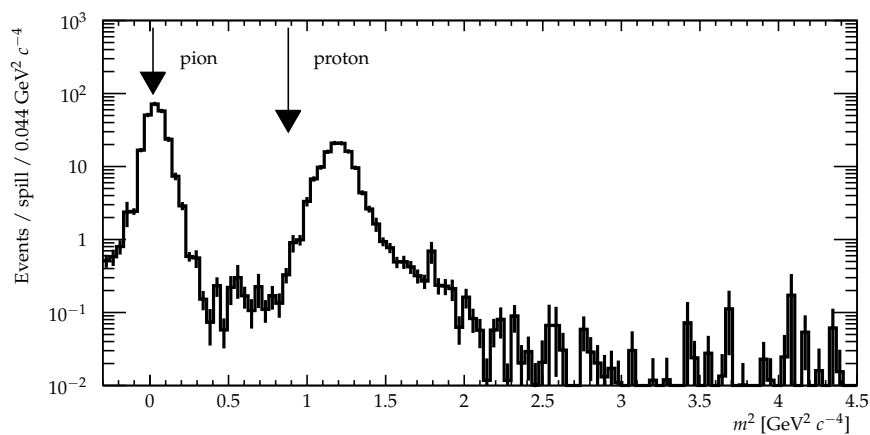}
  \end{adjustbox}
  \caption{Reconstructed mass spectrum for the data taken without moderator blocks. The spectrum was calculated using the time difference between $\mathit{S4}$ and $\mathit{S2}$. Vertical arrows show predicted position of particles.}
  \label{fig:s4tof_mass}
\end{figure}

Additionally, the reconstructed mass distribution for particles travelling from $\mathit{S2}$ to $\mathit{S4}$ is shown in Figure~\ref{fig:s4tof_mass}, produced using Equation~\eqref{eq:recoMass}.
Unlike the same distribution produced for particles travelling from $\mathit{S1}$ to $\mathit{S3}$ (see Figure~\ref{fig:s3tof_mass}), no deuteron peak is visible.
This is thought to be due to the attenuation of deuterons within the walls of the TPC.
Additionally, the predicted proton position does not line up with the measured proton position. 
This is again thought to be caused by the positioning of the TPC in front of $\mathit{S4}$.
Protons passing through the TPC lose energy, resulting in them having less than the original 0.8~GeV/c beam momentum.
In turn, this leads to protons having a larger reconstructed mass than predicted.
The displacement of the proton mass peak in Figure~\ref{fig:s4tof_mass} is consistent with the expected energy loss in the vessel walls.
This consistency is shown with Monte Carlo studies in Section~\ref{sec:mcStudies}.
These Monte Carlo studies also show that, at the energies used in this study, approximately 40\% of protons which impinge on the vessel stop within it.

\begin{figure}[H]
  \begin{adjustbox}{width=.65\textwidth, center}
    \input{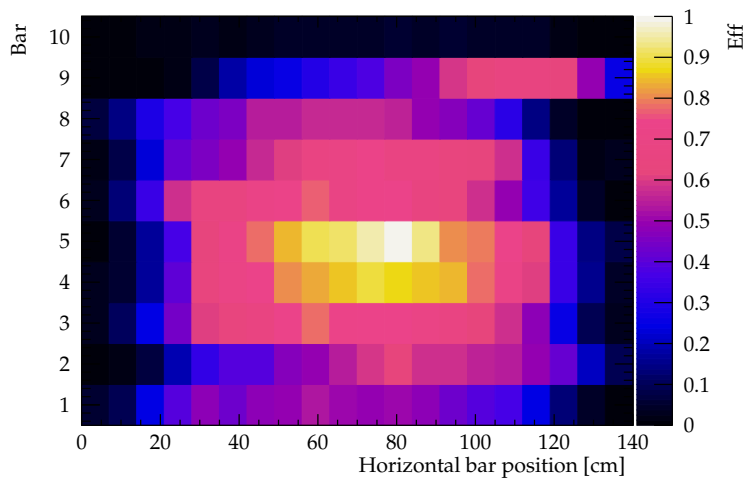}
  \end{adjustbox}
  \caption{Relative detection efficiency of $\mathit{S4}$ as a function of bar number and position along each bar as measured with cosmic rays. The data from bar 10 was not used in the analysis due to the poor efficiency along the bar.}
  \label{fig:s4PosEff}
\end{figure}

A correction is made for the variation in particle detection efficiency between the bars and for the variation in this efficiency as a function of the position along each bar.
This correction is performed using the cosmic ray flux.
It is assumed that the flux of cosmic rays passing through each part of $\mathit{S4}$ is equal.
Each $\mathit{S4}$ bar is divided into 7~cm segments for analysis, and the number of cosmic rays passing through each segment is measured by assuming that all signals occurring outside of beam spills are produced by cosmic rays.
The efficiency is then found from this distribution by normalising the bin with the highest number of cosmic ray signals to 1.
This efficiency is highest around the middle of the bars (70~cm) because of the requirement that coincident signals are observed in both PMTs on a given bar in order for a hit to be recorded.
An example of one of these distributions is shown in Figure~\ref{fig:s4PosEff}.
Events are then weighted according to the bar in which they are observed and their measured position along this bar.
The weight applied is the inverse of the value shown on the z-axis of Figure~\ref{fig:s4PosEff}.
Additionally, a further weight is applied to all $\mathit{S4}$ events of 1.25.
This weight is derived from tests performed on the $\mathit{S4}$ bars with a $^{90}$Sr source.
Using this source, it was determined that the maximum measured rate of signals produced by the $^{90}$Sr source was equal to 0.8 of the true rate.

\vspace{-6pt}

Using Figure~\ref{fig:s4tof}, protons and MIPs are selected with timing cuts and a flat background is then subtracted.
The particles in the quicker timing window (those for which $36~\text{ns}<t_{\mathit{S4}}-t_{\mathit{S2}}<51~\text{ns}$) are considered to be minimum ionizing particles while those in the slower timing window (those for which $62~\text{ns}<t_{\mathit{S4}}-t_{\mathit{S2}}<125~\text{ns}$) are considered to be protons.

\begin{figure}[H]
  \begin{adjustbox}{max totalsize={.75\textwidth}{.62\textheight},center}
    \input{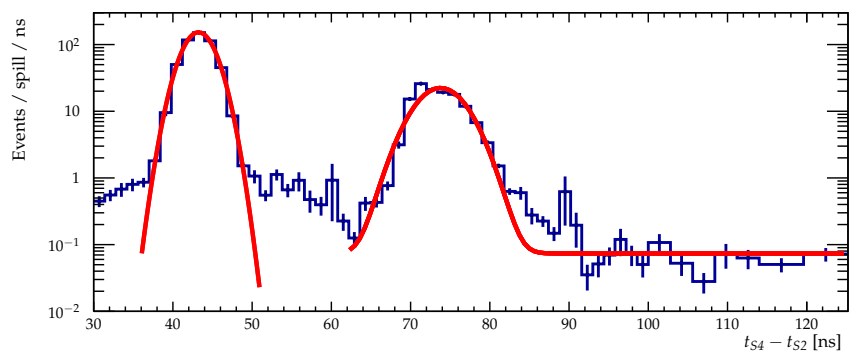}
  \end{adjustbox}
  \caption{Example of the time of flight spectrum observed in $\mathit{S4}$ with combined signal and background functions fitted (shown in red).}
  \label{fig:fitEx}
\end{figure}

The background is determined by fitting a sum of signal and background functions to the time of flight spectra.
The signal functions are taken to be Gaussians while the background is taken to be flat. 
An example of this is shown in Figure~\ref{fig:fitEx}.
The background rates for each sample are shown in Table~\ref{tab:backgrounds}.
These backgrounds have been subtracted from the totals shown in Section~\ref{sec:s4Flux}.

The backgrounds follow the same pattern as the total measured $\mathit{S4}$ particle rates (see Section~\ref{sec:s4Flux}).
The background rate initially increases with the addition of the addition of more moderator blocks then decreases for the 3 and 4 moderator block configurations.

The ratio of the rate of signal protons to the background rate falls with the addition of moderator blocks.
This is due to increased scattering from the moderator blocks which causes more particles to strike $\mathit{S4}$ without passing through $\mathit{S2}$.
This leads to an increase in false coincidences which contribute to the background rate.

\begin{table}[H]
  \centering
  \caption{Background rates for the time of flight spectra measured in $\mathit{S4}$. To convert these to the number of expected background events in a spill, the rate is multiplied by the size of the timing window for either MIPs or protons.}
  \begin{tabular}{cc}
    \toprule
    \textbf{Number of Moderator Blocks} & \textbf{Background / $\text{Events} \times \text{spill}^{-1} \times \text{ns}^{-1}$} \\
    \midrule
    0 & $0.037 \pm 0.004$ \\
    1 & $0.066 \pm 0.005$ \\
    2 & $0.165 \pm 0.007$ \\
    3 & $0.124 \pm 0.009$ \\
    4 & $0.085 \pm 0.002$ \\
    \bottomrule
  \end{tabular}
  \label{tab:backgrounds}
\end{table}

\section{Beam Flux Measurement}
\label{hptpcPaper:sec:Results}
\vspace{-6pt}
\subsection{Flux Measurements with S3}

The ToF systems are at an off-axis angle with respect to the beam axis (see Table~\ref{tab:angS1}), in order to probe the reduced proton momentum spectrum, to cover the region most relevant for neutrino experiments and to measure the flux passing through the TPC.
This is quantified in terms of $\theta$ and $\phi$ (see Section~\ref{sec:coord}).

\begin{figure}[H]
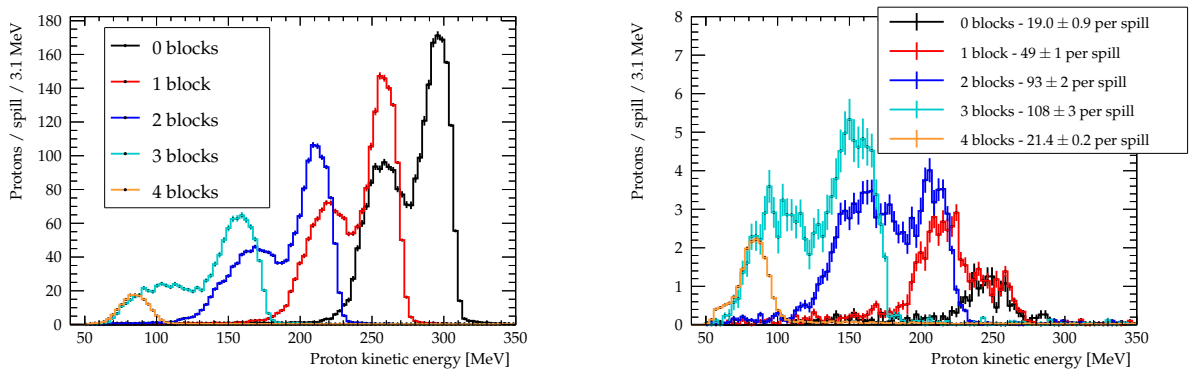

  \centering
  \begin{minipage}[t]{0.49\textwidth}
    \begin{adjustbox}{max totalsize={\textwidth}, center}
      \input{files/Figures/proKeAll.tex}
    \end{adjustbox}
  \end{minipage}
  \hfill
  \begin{minipage}[t]{0.49\textwidth}
    \begin{adjustbox}{max totalsize={\textwidth}, center}
      \input{files/Figures/protonKeTpc.tex}
    \end{adjustbox}
  \end{minipage}
  \caption{\label{fig:s3proke}Proton kinetic energy spectrum as measured in $\mathit{S3}$. (\textbf{Left}) All protons. (\textbf{Right}) The subset of protons passing through the HPTPC drift volume. The errors shown in the legend are the statistical error in particle number per spill.}
\end{figure}

The proton spectra upstream of the TPC are shown in Figure~\ref{fig:s3proke}.
Figure~\ref{fig:s3proke} left, shows the kinetic energy of particles identified as protons is successfully reduced with increasing numbers of moderator blocks, with the range falling from 210-320~MeV for the unmoderated beam, to 60--110~MeV for 4~acrylic blocks.
Figure~\ref{fig:s3proke} right, shows the kinetic energy spectrum of protons crossing into the TPC.
This~figure indicates that the flux of low energy protons (those with a kinetic energy of less than 80~MeV) reaching the TPC was increased from negligible in the 0, 1 and 2 block cases to ($9.7 \pm 0.1$) per spill for the 4 block case.
Comparing Figure~\ref{fig:s3proke} right, with Figure~\ref{fig:protonsfromargon} shows that, for the four moderator block case, the kinetic energy of protons incident upon the TPC is just above the 50~MeV region where the different neutrino interaction generators become discrepant.
These protons lose further energy within the walls of the HPTPC vessel, resulting in a flux of protons below 50~MeV within the TPC.

The combination of the use of moderator blocks and positioning the TPC in an off-axis position also caused a change in the multiplicity of protons passing through the TPC.
Figure~\ref{fig:s3proke} right, shows that the addition of 1, 2 and 3 moderator blocks increased the number of protons passing through the TPC from ($19.0 \pm 0.9$) per spill in the unmoderated case to ($108 \pm 3$) per spill in the 3 block case.
The addition\mbox{} of the fourth moderator block effectively removes the flux of protons above 100~MeV, leaving ($21.4 \pm 0.2$) per spill to traverse the TPC active volume.
  
\begin{figure}[h]
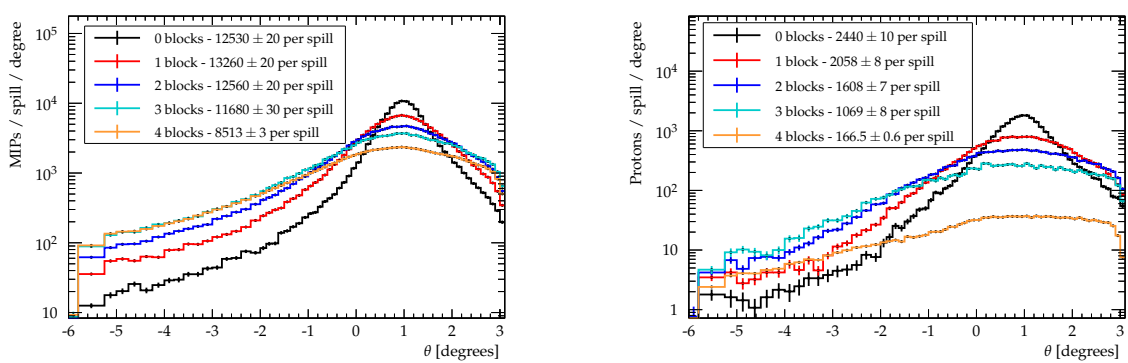

  \begin{minipage}{0.48\textwidth}
    \begin{adjustbox}{max totalsize={\textwidth}{.35\textheight},center}
      \input{files/Figures/thetaS1pi.tex}
    \end{adjustbox}
  \end{minipage}
  \hspace{0.3cm}
  \begin{minipage}{0.48\textwidth}
    \begin{adjustbox}{max totalsize={\textwidth}{.35\textheight},center}
      \input{files/Figures/thetaS1pro.tex}
    \end{adjustbox}
  \end{minipage}
    \caption{\label{fig:s1s3mips}Distribution of hits in $\mathit{S3}$ as a function of the horizontal off-axis angle, measured from $\mathit{S1}$, for varying numbers of moderator blocks. No coincident hit in $\mathit{S2}$ was required. (\textbf{Left}) Minimum ionizing particles. (\textbf{Right}) Protons. The errors shown in the legend are the statistical error in particle number per~spill.}
\end{figure}

  The distributions vs. off-axis angle of MIPs and protons in $\mathit{S3}$ are shown in Figure~\ref{fig:s1s3mips}.
In both cases, the peak beam intensity falls and broadens in $\theta$ with the increasing number of moderator blocks. 
At off-axis angles the number of MIPs and protons is increased as the number of moderator blocks is increased.
The TPC lies within this off-axis region.
The spread of particles for unmoderated data was unexpected; this peak was broadened by the beam steering scattering that led to the double proton peaks seen in Figures~\ref{fig:s3tof} and \ref{fig:s3tof_mass}.
For the unsteered and unmoderated beam, the measured horizontal FWHM is 9.6~cm while the vertical FWHM is 11.0~cm.
This is compared with the measured horizontal FWHM for the unmoderated and steered beam of 16.8~cm.

\begin{figure}[h]
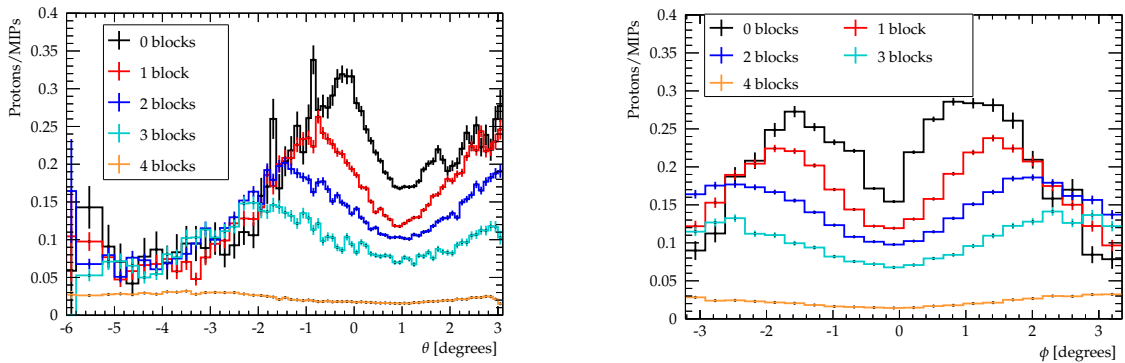

  \begin{minipage}[t]{0.48\textwidth}
    \begin{adjustbox}{max totalsize={\textwidth}{.35\textheight}, center}
      \input{files/Figures/thetaS1ratio.tex}
    \end{adjustbox}
    
  \end{minipage}
  \hspace{0.3cm}
  \begin{minipage}[t]{0.48\textwidth}
    \begin{adjustbox}{max totalsize={\textwidth}{.35\textheight}, center}
      \input{files/Figures/phiS1ratio.tex}
    \end{adjustbox}
  \end{minipage}	
  \caption{\label{fig:propiratio_s3_horz}Proton--MIP ratio in $\mathit{S3}$ for varying numbers of moderator blocks as a function of off-axis angle, as measured from $\mathit{S1}$.  (\textbf{Left}) Horizontal angle. (\textbf{Right}) Vertical angle. The TPC spans horizontal angles 1.4--3.6$^{ \circ }$ and vertical angles $-2.6$--$+2.6$$^{ \circ }$.}
\end{figure}

Figure~\ref{fig:propiratio_s3_horz} shows the proton--MIP ratio measured in $\mathit{S3}$ as a function of the nominal off-axis angle, horizontally and vertically, respectively, and for various numbers of moderator blocks.
For 0, 1, 2 and 3 moderator blocks the ratio falls to a minimum at approximately $1^{\circ}$ with respect to the beam axis.
This corresponds to the true beam centre for the steered beam.
As the angle moves away from the true beam centre, the ratio rises for these configurations.
The peak of the proton--MIP ratio shifts away from the beam centre progressively as more moderator blocks are added (from approximately $1^{\circ}$ away from beam centre for 0 blocks up to approximately $3^{\circ}$ away from beam centre for 3 blocks).
At most values of $\theta$, the proton--MIP ratio falls with the addition of more moderator blocks. 
Thus, reducing the kinetic energy of the protons below 100~MeV came at the cost of reducing the purity of the proton beam.

\subsection{Flux Measurements with S4}
\label{sec:s4Flux}

\begin{figure}[h]
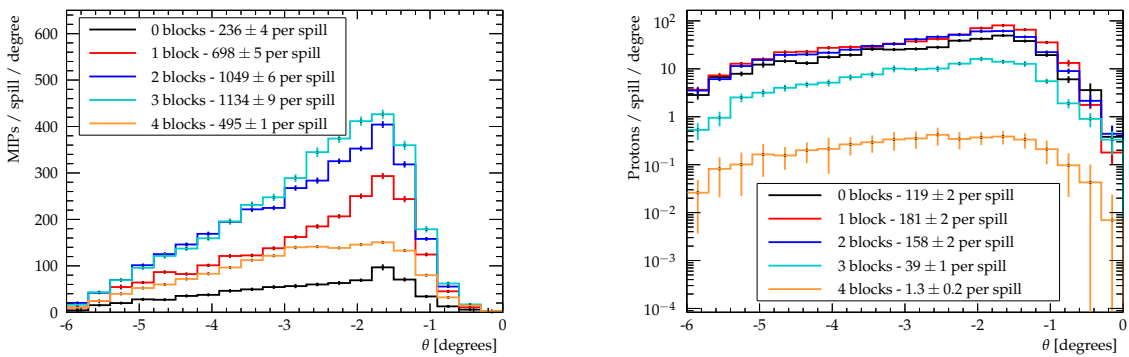

  \begin{minipage}[t]{0.48\textwidth}
    \begin{adjustbox}{max totalsize={\textwidth}{.35\textheight},center}
      \input{files/Figures/piS4Horz_new.tex}
    \end{adjustbox}
    
  \end{minipage}
  \hspace{0.3cm}
  \begin{minipage}[t]{0.48\textwidth}
    \begin{adjustbox}{max totalsize={\textwidth}{.35\textheight},center}
      \input{files/Figures/proS4Horz_new.tex}
    \end{adjustbox}
  \end{minipage} 
  \caption{\label{fig:thetas4mip}Distribution of hits in $\mathit{S4}$ as a function of the number of moderator blocks and the horizontal off-axis angle. (\textbf{Left}) Minimum ionizing particles. (\textbf{Right}) Protons. The errors shown in the legend are the statistical error in particle number per spill.}
\end{figure}

Figure~\ref{fig:thetas4mip} left, shows the flux of particles identified as minimum ionizing particles across $\mathit{S4}$.
For all numbers of moderator blocks, the peak number of minimum ionizing particle events occurs at a value of $\theta$ between $-1^{\circ}$ and $-2^{\circ}$.
Similarly the number of proton events per spill, shown in Figure~\ref{fig:thetas4mip} right, peaks at a value of $\theta$ of approximately $-2^{\circ}$.
The fall in the number of events between $\theta = -1^{\circ}$ and $\theta = 0^{\circ}$ is as a result of the beam impinging on the vessel doors at these angles.
The positioning and shape of the pressure vessel doors means that, for particles travelling at these angles, a greater length of steel is passed through compared to those particles which strike the body of the vessel.

Figure~\ref{fig:thetas4mip} left, also shows that initially, an increasing number of moderator blocks results in an increased total MIP flux through $\mathit{S4}$. 
This is because both $\mathit{S2}$ and $\mathit{S4}$ are positioned off-axis, so the unmoderated beam particles do not strike these detectors.
Due to scattering processes in the moderator, a greater number of MIPs are incident upon $\mathit{S2}$ and $\mathit{S4}$, with more scattering occurring with greater numbers of moderator blocks.
However, with the fourth moderator block the flux of MIPs is seen to fall.
Similarly, with the addition of the first two moderator blocks, the proton flux shown in Figure~\ref{fig:thetas4mip} right, initially sees an increase in the total number of events in $\mathit{S4}$.
However, with three and four moderator blocks, the total number of protons observed in $\mathit{S4}$ falls.
The initial proton flux increase is similar to that for the MIP flux, with increased scattering causing more protons to pass through the off-axis $\mathit{S2}$ and $\mathit{S4}$ detectors.
The subsequent decrease is due to the larger loss of energy of the protons in the thicker moderator.
In turn, this leads to attenuation of protons in the pressure vessel resulting in fewer observed events in $\mathit{S4}$.

\begin{figure}[h]
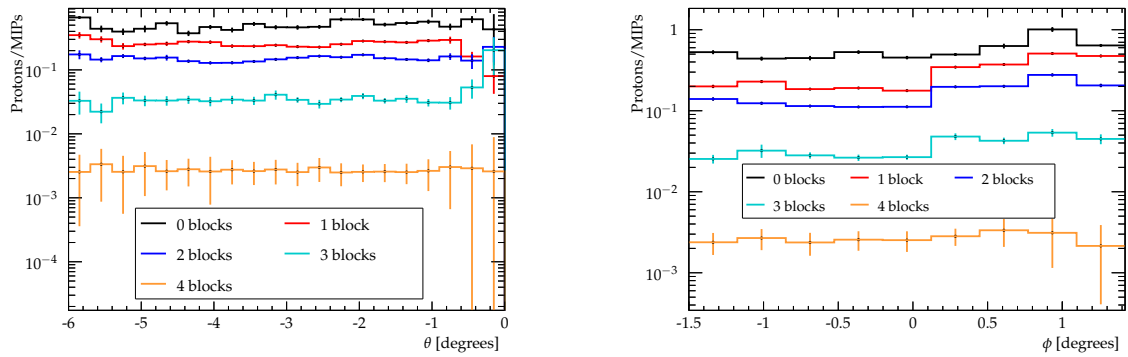

  \begin{minipage}[t]{0.48\textwidth}
    \begin{adjustbox}{max totalsize={\textwidth}{.35\textheight}, center}
      \input{files/Figures/ratioS4Horz_new.tex}
    \end{adjustbox}
  \end{minipage}
  \hspace{0.3cm}
  \begin{minipage}[t]{0.48\textwidth}
    \begin{adjustbox}{max totalsize={\textwidth}{.35\textheight}, center}
      \input{files/Figures/ratioS4Vert_new.tex}
    \end{adjustbox}
  \end{minipage}	
  \caption{\label{fig:propiratio_s4_horz}Proton--MIP ratio in $\mathit{S4}$ for varying numbers of moderator blocks as a function of off-axis angle. (\textbf{Left}) Horizontal off-axis angle. (\textbf{Right}) Vertical off-axis angle.}
\end{figure}

Figure~\ref{fig:propiratio_s4_horz} shows the ratio of protons to MIPs as a function of the number of moderator blocks, $\theta$ and $\phi$.
For all of the different block configurations, the ratio is flat across both $\theta$ and $\phi$.
With the addition of moderator blocks, the ratio reduces from its highest level of 0.5 for the 0 block case, to 0.002 for the 4 block data.
As mentioned previously, this is thought to be due to the attenuation of low energy protons within the walls of the pressure vessel.

\subsection{Monte Carlo Studies}
\label{sec:mcStudies}
In order to ascertain the flux of protons reaching the active region of the TPC, and verify the corrections described above, a Monte Carlo (MC) simulation study was performed.
The simulation was performed using GEANT4~\cite{brun1993geant}, with geometric volumes approximating the vessel, TPC and time of flight systems.
In order to match upstream conditions as closely as possible, particle momenta were drawn from the $\mathit{S3}$ distributions shown in Figure~\ref{fig:s3proke} left, and simulated with trajectories that resulted in the same position distribution as seen in Figure~\ref{fig:s3XY_pion} right.
The same timing cuts described in Section~\ref{timeofflightanalysissubsec} were applied.

\begin{figure}[H]
  \centering
     \begin{adjustbox}{max totalsize={\textwidth}{.3\textheight}, center}
      \input{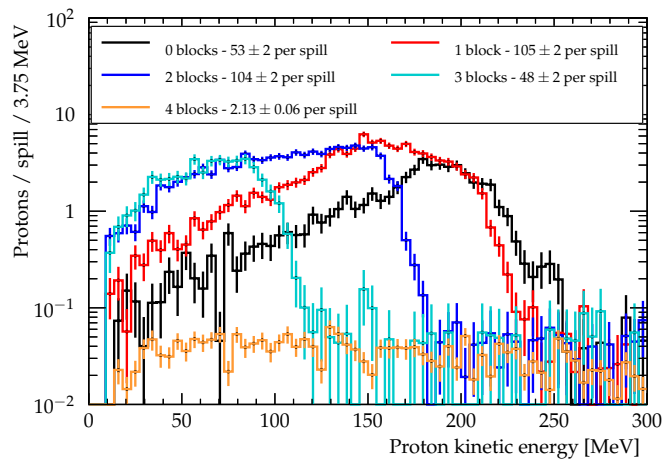}
    \end{adjustbox}
    \caption{Energy profile of simulated protons reaching $\mathit{S4}$, with kinetic energy above the detection threshold of 10~MeV.}
    \label{fig:MCS4}
\end{figure}

The simulated protons are propagated through the vessel to the $\mathit{S4}$ detector.
The momentum profile of simulated protons reaching the $\mathit{S4}$ panel is shown in Figure~\ref{fig:MCS4}.
A proton detection threshold of $\mathit{S4}$ of 140~MeV/c (10~MeV kinetic energy) is included.
The simulation shows a significant reduction in kinetic energy as most particles have travelled through both steel walls of the TPC vessel.
In~particular, in the 4 moderator block case, very few particles have survived through the second vessel wall to reach~$\mathit{S4}$.

\begin{figure}[H]
  \centering
  \begin{adjustbox}{max totalsize={.35\textheight}, center}
    \input{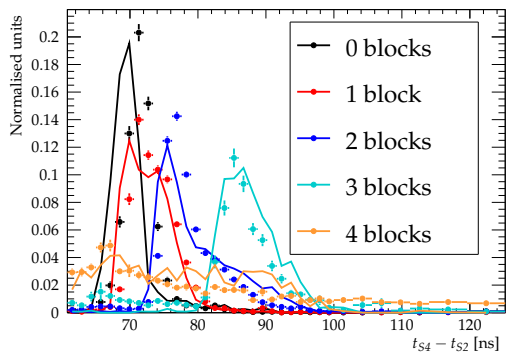}
  \end{adjustbox}
  \caption{Comparison of simulated and measured proton ToF between $\mathit{S2}$ and $\mathit{S4}$. Solid lines correspond to the simulated distributions, while points correspond to data. All distributions are area normalised to 1.}
  \label{fig:tofMC}
\end{figure}

Comparisons of $\mathit{S2}$ to $\mathit{S4}$ time of flight for data and MC are shown in Figure~\ref{fig:tofMC} for varying numbers of moderator blocks.
Figure~\ref{fig:tofMC} shows that, for all numbers of moderator blocks, the peak positions in the data and MC spectra agree to within 2~ns.
This level of agreement confirms that the simulated energy loss in the vessel and TPC is similar to the energy loss in the data.

Systematic uncertainties on the number of protons measured in $\mathit{S3}$ and $\mathit{S4}$ are estimated for both data and MC, and shown in Table~\ref{tab:systematics}.
The systematic uncertainty on the MC simulation is determined by varying the geometric initial conditions of the simulation, including the position of the $\mathit{S1}$ and $\mathit{S2}$ detectors.
These variations induce changes in the direction and momenta of the propagated protons.
Additionally, a study was performed with 1~cm of additional acrylic in the beamline, as a proxy for the uncertainty on other pieces of light material in the beam facility.
This set of calculated errors represents geometric sources of uncertainty in the MC simulation.

For the data, the uncertainty on the overall efficiency of $\mathit{S3}$ is calculated by taking the $\pm1\sigma$ uncertainty on the fitted linear relationship between $\mathit{S1} \cap \mathit{S2}$ signals in the upstream and downstream ToF DAQs (see Section~\ref{timeofflightanalysissubsec}) and calculating the fractional change this causes in the $\mathit{S3}$ proton count.

The uncertainty on the overall efficiency of $\mathit{S4}$ is calculated from the calibration tests performed on the $\mathit{S4}$ bars with a $^{90}$Sr source, as discussed in Section~\ref{timeofflightanalysissubsec}.
The overall efficiency factor of 0.8 was calculated using data taken with a significantly different readout to that used in the beam test and therefore is subject to variation. 
The spread in maximum bar efficiencies measured in these $^{90}$Sr source tests for the various $\mathit{S4}$ bars is used as the systematic uncertainty on the overall $\mathit{S4}$ efficiency.

The $\mathit{S4}$ angular correction systematic uncertainty is assessed by varying the number of horizontal bins in Figure~\ref{fig:s4PosEff} from 20 to 10 and taking the fractional change in the number of measured $\mathit{S4}$ protons.

The uncertainty on the $\mathit{S4}$ background subtraction is determined by taking the $1\sigma$ error on the fitted flat background and determining the resulting change in the number of protons.
This has a larger effect in the 4 block case because of the very small number of protons detected in $\mathit{S4}$ relative to the background.

\begin{table}[H]
  \centering
  \caption{List of systematic errors and values for data and Monte Carlo (MC) simulation. All values are the percent error on the $\mathit{S4}$ proton count with the exception of the uncertainty on the efficiency of $\mathit{S3}$, which is the percent error on the $\mathit{S3}$ proton count. All uncertainties are treated as uncorrelated. $n_{\mathit{S4},~\text{MC}}$ refers to the number of protons reaching $\mathit{S4}$ in MC simulations.}
  \begin{tabular}{cccccc}
    \toprule
    \multicolumn{6}{c}{\textbf{Monte Carlo}} \\
    \midrule
    & \multicolumn{5}{c}{Number of moderator blocks} \\
    & 0 & 1 & 2 & 3 & 4 \\
    \midrule
    Systematic uncertainty on $n_{\mathit{S4},~\text{MC}}$ & 9.5\% & 8.0\% & 8.5\% & 17.0\% & 8.0\% \\
    \midrule
    \multicolumn{6}{c}{\textbf{Data}} \\
    \midrule
    & \multicolumn{5}{c}{Number of moderator blocks} \\
    Source of systematic error & 0 & 1 & 2 & 3 & 4 \\
    \midrule
    Absolute efficiency of $\mathit{S3}$ & 1.1\% & 11.4\% & 7.0\% & 11.4\% & 4.9\% \\
    Absolute efficiency of $\mathit{S4}$ & 11.0\% & 11.0\% & 11.0\% & 11.0\% & 11.0\% \\ 
    $\mathit{S4}$ angular correction & 2.9\% & 1.5\% & 6.7\% & 8.2\% & 4.1\% \\
    $\mathit{S4}$ background uncertainty & 0.18\% & 0.16\% & 1.1\% & 1.4\% & 8.1\% \\
    \midrule
    \textbf{Total} & 11.5\% & 16.0\% & 14.7\% & 18.3\% & 18.9\% \\
    \bottomrule 
  \end{tabular}
  \label{tab:systematics}
\end{table}

The ratio of number of protons reaching $\mathit{S4}$ to those reaching $\mathit{S3}$ is shown for both simulation and data in Table~\ref{tab:ratios}, which includes the total statistical and systematic error in each case.
The agreement shown relative to the uncertainty provided by the beam test setup provides strong evidence that efficiency corrections described in Section~\ref{timeofflightanalysissubsec} are justified.

\begin{figure}[H]
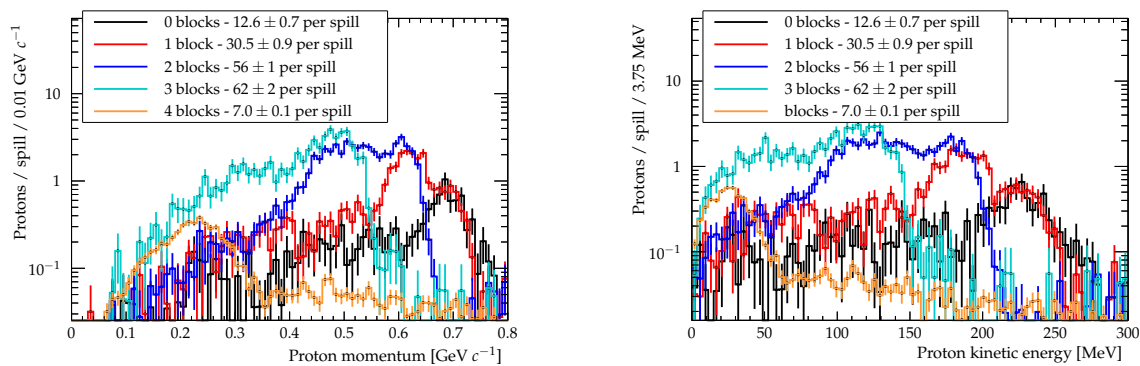

  \begin{minipage}[t]{0.48\textwidth}
    \begin{adjustbox}{max totalsize={\textwidth}{.35\textheight}, center}
      \input{files/Figures/momTpc.tex}
    \end{adjustbox}
  \end{minipage}
  \hspace{0.3cm}
  \begin{minipage}[t]{0.48\textwidth}
    \begin{adjustbox}{max totalsize={\textwidth}{.35\textheight}, center}
      \input{files/Figures/keTpc.tex}
    \end{adjustbox}
    
  \end{minipage}
  \caption{(\textbf{Left}) Momentum profile of simulated protons reaching the active region of the TPC. (\textbf{Right}) Energy profile of simulated protons reaching the active region of the TPC. The errors shown in the legend are the statistical error in particle number per spill.}
  \label{fig:MCTPC}	
\end{figure}

The number of simulated particles that penetrate the active area of the TPC are shown in Figure~\ref{fig:MCTPC} left and right, as a function of momentum and kinetic energy, respectively.
Comparing Figure~\ref{fig:MCTPC} right, with the motivation plot shown in Figure~\ref{fig:protonsfromargon}, it is clear that 4 moderator blocks were required to access the momentum region of interest (below 50 MeV).
The off-axis and moderator technique were therefore successful in the extent to which the proton energy was lowered.
The number of protons reaching the active area of the TPC was  per spill ($7.0 \pm  0.1$) for 4 moderator blocks, compared with ($12.6 \pm 0.7$) per spill without moderation. 
For 4 moderator blocks, ($5.6 \pm  0.1$) of those protons had energies below 100~MeV.
These values were calibrated with the full comparison between data and~simulation.

\begin{table}[H]
  \centering
  \caption{Ratio of number of protons reaching $\mathit{S4}$ to number protons reaching $\mathit{S3}$ for different numbers of moderator blocks in MC and data. In each instance, the combined statistical and systematic errors are shown.}
  \begin{tabular}{cccc}
   \toprule
    \textbf{Number of Moderator Blocks} & \textbf{Monte Carlo} & \textbf{Data} & \textbf{Data/MC}\\
    \midrule
    $0$ & $0.027 \pm 0.003$ & $0.049 \pm 0.007$ & $1.8 \pm 0.3$ \\
    $1$ & $0.067 \pm 0.005$ & $0.09 \pm 0.01$ & $1.3 \pm 0.2$ \\
    $2$ & $0.084 \pm 0.007$ & $0.10 \pm 0.01$ & $1.2 \pm 0.2$ \\
    $3$ & $0.06 \pm 0.01$ & $0.036 \pm 0.007$ & $0.7 \pm 0.2$ \\
    $4$ & $0.011 \pm 0.001$ & $0.008 \pm 0.001$ & $0.7 \pm 0.1$ \\
    \bottomrule
  \end{tabular}
  \label{tab:ratios}
\end{table}

       \section{Conclusions}
\label{hptpcPaper:sec:Conclusion}

The prototype high pressure gas time projection chamber was operated in the T10 beamline at CERN in August and September 2018 in order to make measurement of low momentum protons in argon.
The vessel was placed at a position off the centre axis of the beam, and a number of acrylic blocks were placed directly in the beamline in order to produce a flux of low momentum protons through the TPC, ensure a low occupancy of these low energy protons within the TPC and change the ratio of MIPs to protons.
Measurements of the beam flux were made using two time of flight systems placed ($1.323 \pm 0.001$)~m upstream and ($0.918 \pm 0.001$)~m downstream of the TPC vessel.
\mbox{These measurements} were used to determine the absolute and relative rates of protons and MIPs as well as their momenta, at different positions off the beam axis, and for varying numbers of moderator blocks.

When the beam was unsteered, the width was measured to be 9.6~cm.
When the beam was steered approximately $1^{\circ}$ off-axis, the beam width increased to 16.8~cm.

These measurements demonstrated that adding moderator blocks reduced the average kinetic energy of protons reaching the TPC from 0.3~GeV with 0~moderator blocks to 0.1~GeV for 4 moderator blocks, accessing the kinematic region of interest.
This indicates that the off-axis moderator technique provides a suitable method for providing low energy hadron beams for neutrino detector tests.
\mbox{The proton/MIP} ratio increased at low off-axis angles, peaking at 1$^{ \circ }$–2$^{ \circ }$ off axis, depending on how many moderator blocks were used and then fell off at higher angles.
The four moderator block configuration yielded a proton/MIP ratio that was  substantially lower than 0–3 blocks and also flat versus off-axis angle, but achieved the desired proton energy spectrum.
With calibration from the upstream and downstream time of flight systems, for data with 4 moderator blocks in the beamline the simulated number of protons with energy below 100~MeV reaching the active TPC region was ($5.6 \pm  0.1$) per spill with an energy range of 0 to 50~MeV/c.

\authorcontributions{{Conceptualization, A.B.-F., S.B., L.C., J.H., A.K. (Asher Kaboth), J.M., R.N., J.N., R.S., N.S., Y.S. and M.W. (Morgan Wascko); data curation, S.J., T.N., E.A., D.B., A.D. (Alexander Deisting), A.D. (Adriana Dias), P.D., J.H., P.H.-B., A.K. (Asher Kaboth), A.K. (Alexander Korzenev), M.M., J.M., R.N., J.N., W.P., H.R.-Y., Y.S., A.T., M.U., S.V., A.W. and M.W. (Morgan Wascko); formal analysis, S.J., T.N., P.D., A.K. (Alexander Korzenev), Y.S. and A.W.; funding acquisition, G.B., A.B.-F., S.B., A.K. (Asher Kaboth), J.M., R.N., J.N., S.R., R.S. and M.W. (Morgan Wascko); investigation, S.J., T.N., A.B.-F., S.B., L.C., P.D., P.H.-B., A.K. (Asher Kaboth), W.M., M.M., J.M., R.N., J.N., R.S., Y.S., J.S., M.U., S.V., A.W., M.W. (Mark Ward) and M.W. (Morgan Wascko); methodology, S.J., T.N., E.A., A.B.-F., C.B., S.B., Z.C.-W., L.C., A.D. (Alexander Deisting), A.D. (Adriana Dias), P.D., J.H., A.K. (Asher Kaboth), A.K. (Alexander Korzenev), P.M., J.M., R.N., W.P., H.R.-Y., R.S., N.S., Y.S., A.T., M.U., S.V., A.W., M.W. (Mark Ward) and M.W. (Morgan Wascko); project administration, G.B., S.B., L.C., A.D. (Alexander Deisting), A.K. (Asher Kaboth), J.M., R.N., J.N. and M.W. (Morgan Wascko); resources, S.B., L.C., A.K. (Asher Kaboth), J.M., R.N., J.N., S.R. and M.W. (Morgan Wascko); Software, S.J., T.N., E.A., D.B., Z.C.-W., L.C., A.D. (Alexander Deisting), A.D. (Adriana Dias), J.H., A.K. (Asher Kaboth), J.M., R.N., W.P., S.V., A.W. and M.W. (Mark Ward); supervision, G.B., A.B.-F., S.B., L.C., A.D. (Alexander Deisting), P.D., J.H., A.K. (Asher Kaboth), J.M., R.N., J.N., R.S., A.W. and M.W. (Morgan Wascko); validation, S.J., T.N., E.A., D.B., Z.C.-W., L.C., A.D. (Alexander Deisting), A.D. (Adriana Dias), A.K. (Asher Kaboth), M.M., J.M., R.N., W.P., H.R.-Y. and A.T.; visualization, S.J., T.N., L.C. and R.N.; writing---original draft, S.J., T.N., A.K. (Asher Kaboth), J.M. and M.W. (Morgan Wascko); writing---review and editing, S.J., T.N., E.A., S.B., D.B., Z.C.-W., L.C., A.D. (Alexander Deisting), A.D. (Adriana Dias), P.D., A.K. (Asher Kaboth), M.M., J.M., R.N., J.N., W.P., H.R.-Y., A.T., A.W. and M.W. (Morgan Wascko).}
 All authors have read and agreed to the published version of the manuscript.}

\funding{{This research was funded in part by Science and Technology Facilities Council grant number ST/N003233/.} 
}

\acknowledgments{We wish to acknowledge support for summer students from the Ogden Trust and St. Andrews University, and outstanding support during the beam test from Johannes Bernhard of CERN as well as
		Rebecca Conybeare, Nicole Cullen, Kate Gould, Veera Mikola, Christopher Thorpe, and	Simon Williams.}

\conflictsofinterest{{The authors declare no conflict of interest.} 
}

\reftitle{References}

\end{document}